\documentclass[a4paper,12pt]{elsarticle}

\usepackage{tikz}
\usepackage{amsmath}

\usepackage{filecontents}

\usepackage{lineno,hyperref}
\usepackage{inconsolata}
\usepackage{url}
\usepackage{amsmath}
\usepackage{etoolbox}
\usepackage{graphicx}
\usepackage{listings}
\usepackage{listings-rust}
\usepackage{subcaption}
\usepackage{cleveref}
\usepackage[export]{adjustbox}
\usepackage{float}
\usepackage{xcolor}
\PassOptionsToPackage{dvipsnames,xcdraw}{xcolor}
\usepackage{soul}
\usepackage{booktabs}

\usepackage{todo}

\newcommand{\systemname}{UPSS}
\newcommand{\systemfullname}{UPSS: the user-centric private sharing system}
\newcommand{\fsname}{UPSS-FUSE}
\newcommand{\vcname}{UVC}
\newcommand{\vcfullname}{UVC: UPSS Version Control System}

\definecolor{codegreen}{rgb}{0,0.6,0}
\definecolor{codegray}{rgb}{0.5,0.5,0.5}
\definecolor{codepurple}{rgb}{0.58,0,0.82}
\definecolor{backcolour}{rgb}{0.95,0.95,0.92}

\lstdefinelanguage{diff}{
  morecomment=[f][\color{codepurple}]{@@},
  morecomment=[f][\color{codegreen}]{+++\ },
  morecomment=[f][\color{codegreen}]{+\ },
  morecomment=[f][\color{red}]{---\ },
  morecomment=[f][\color{red}]{-\ },
}

\newcommand{\ccode}[1]{\lstinline[basicstyle=\small\ttfamily,language=C]{#1}}
\newcommand{\command}[1]{\lstinline[basicstyle=\ttfamily]{#1}}
\newcommand{\data}[1]{\lstinline[basicstyle=\small\ttfamily]{#1}}
\newcommand{\object}[1]{\lstinline[basicstyle=\ttfamily]{#1}}
\newcommand{\rust}[1]{\lstinline[basicstyle=\ttfamily,language=Rust]{#1}}

\begin{document}
\begin{frontmatter}
\title{\systemname{}: a User-centric Private Storage System with its applications}

\author[one]{Arastoo~Bozorgi\corref{cor1}} 
\ead{ab1502@mun.ca}
\author[one]{Mahya~Soleimani~Jadidi}
\ead{msoleimanija@mun.ca}
\author[one]{Jonathan~Anderson}
\ead{jonathan.anderson@mun.ca}
\affiliation[one]{organization={Department of Electrical and Computer Engineering},
            addressline={Memorial University},
            city={St. John's},
            postcode={},
            state={NL},
            country={Canada}}
\cortext[cor1]{Corresponding author}

\begin{abstract}
	Strong confidentiality, integrity, user control,
	reliability and performance are critical requirements in privacy-sensitive applications.
	Such applications would benefit from a data storage and sharing
	infrastructure that provides these properties even in decentralized topologies
	with untrusted storage backends, but users today are forced to choose between
	systemic security properties and system reliability or performance.
	As an alternative to this \textit{status quo} we present
	\textit{\systemfullname{}}, a cryptographic storage system that can be used as a
	conventional filesystem or as the foundation for security-sensitive applications
	such as redaction with integrity and private revision control.
	We demonstrate that both the security and performance properties of
	\systemname{} exceed that of existing cryptographic filesystems and that its performance is comparable
  to mature conventional filesystems --- in some cases even superior.
  Whether used directly via its Rust API or as a conventional filesystem,
  \systemname{} provides strong security and practical performance on untrusted storage.
\end{abstract}

\begin{keyword}
	Cryptographic filesystem, distributed filesystem, private sharing, redaction, private version control.
\end{keyword}
\end{frontmatter}


\section{Introduction}
\label{sec:introduction}
Across a broad spectrum of domains, there is an acute need for
private storage with flexible, granular sharing.
Environments as diverse as social networking,
electronic health records and surveillance data management require both
strong cryptographic protection and fine-grained sharing across
security boundaries without granting overly-broad access.
Existing systems provide coarse security guarantees or strong performance
properties, but rarely both.
Fine-grained, flexible, high-performance sharing of default-private data
is still a challenging problem.

What is needed is a mechanism for
\textit{least-privileged} storage that facilitates
\textit{simple discretionary sharing} of arbitrary subsets of data,
providing strong confidentiality and integrity properties
on commodity cloud services from untrusted providers.
In the previous years, some cryptographic filesystems 
have been developed that store user data on untrusted storage 
providers. However, they cannot provide strong security properties nor
flexible data sharing. For example, EncFS \cite{encfs} and CryFS \cite{messmer2017novel}
are cryptographic filesystems that leave metadata unprotected, or in the latter
one, everything is encrypted with one key. TahoeFS \cite{wilcox2008tahoe} is
another cryptographic filesystem with strong security properties, but its design
does not allow flexible and fine-grained data sharing.

In this paper, we have built \systemfullname{}, which 
is a ``global first'' cryptographic filesystem with no
assumptions of trustworthiness for storage infrastructure
or even on common definitions of user identities.
Relying on key concepts from capability systems~\cite{capabilitysystems},
distributed systems, log-structured filesystems
and revision control, we have developed a new approach to
filesystems that offers novel features while being usable
in ways that are compatible with existing applications.

\systemname{} makes several key contributions to the field of
privacy-preserving filesystems. First, unlike 
cryptographic filesystems that entangle user 
and group identifications and device specification with access controls, 
\systemname{} stores 
all data as encrypted blocks on untrusted block stores including 
local, network, or cloud block stores, without any mapping between the 
blocks or blocks to block owners. 
Granular access controls are then defined by higher level 
applications according to application semantics. 
Traditional access control modalities such as Unix permissions can be 
implemented by systems using \systemname{}, as in the case of our FUSE-based interface, 
but they are not encoded in the shared cryptographic filesystem itself. 
This decoupling allows the filesystem to be global-first and local-second.

Second, all \systemname{} blocks can be accessed by cryptographic capabilities~\cite{capabilitysystems}
called block pointers consist of block names and their decryption keys
that reduces the burden of key management and simplifies naming; a block pointer
is enough to fetch, decrypt and read a block, with no central key management required.
Block pointers enable flexible data sharing at the block level among
mutually-distrustful users. They also enable per-block encryption rather
than per-file or per-filesystem encryption, which provides a stronger security model.



Third, \systemname{} enables aggressive and safe caching by defining
a multi-layer caching block store consists of other block stores that
guarantee data consistency between all block stores.
The caching block store prioritizes applying the operations on faster
block stores on the cache hierarchy and processes the operations on 
slower block stores in the background. Therefore, the caching block 
store becomes available immediately despite the number of layers in
the hierarchy or the slowness of higher-level block stores.
This provides performance that exceeds cryptographic 
filesystems by factors of 1.5–40$\times$.

Finally, \systemname{} design enables novel security and privacy operations such
as provenance-preserving redaction and private-by-default revision control.

\systemname{}' system model and design is described in \Cref{sec:upss}.
Its security model is described in \Cref{sec:security-model},
with specific comparison to the security properties of both conventional and
cryptographic filesystems.
Performance is evaluated in \Cref{sec:eval} via three case studies
comparing \systemname{} to existing filesystems:
local filesystems (\Cref{sec:local-fuse}),
network filesystems (\Cref{sec:network-fuse}) and
global filesystems (\Cref{sec:global-fuse}).
Finally, novel applications enabled by \systemname{} are explored in
\Cref{sec:applications},
including provenance-preserving redaction (\Cref{sec:redaction})
and a new model of private revision control (\Cref{sec:vcs}).

\section{\systemname{} system model and design}
\label{sec:upss}
\systemname{} is a \textit{cloud-first} private storage and sharing system.
Rather than a local cryptographic filesystem that projects POSIX assumptions
(e.g., file ownership, user identification and trusted devices)
into the cloud, \systemname{} starts with the assumptions of untrusted storage
and user-directed sharing via
cryptographic capabilities~\cite{capabilitysystems}.
\systemname{} can be exposed via FUSE~\cite{fuse} as a conventional
POSIX filesystem, allowing performance comparison to existing
local filesystems, network filesystems and global filesystems,
but its most exciting capabilities are exposed directly through a
Rust API.

\begin{figure}
	\centering
	\includegraphics[width=0.9\columnwidth]{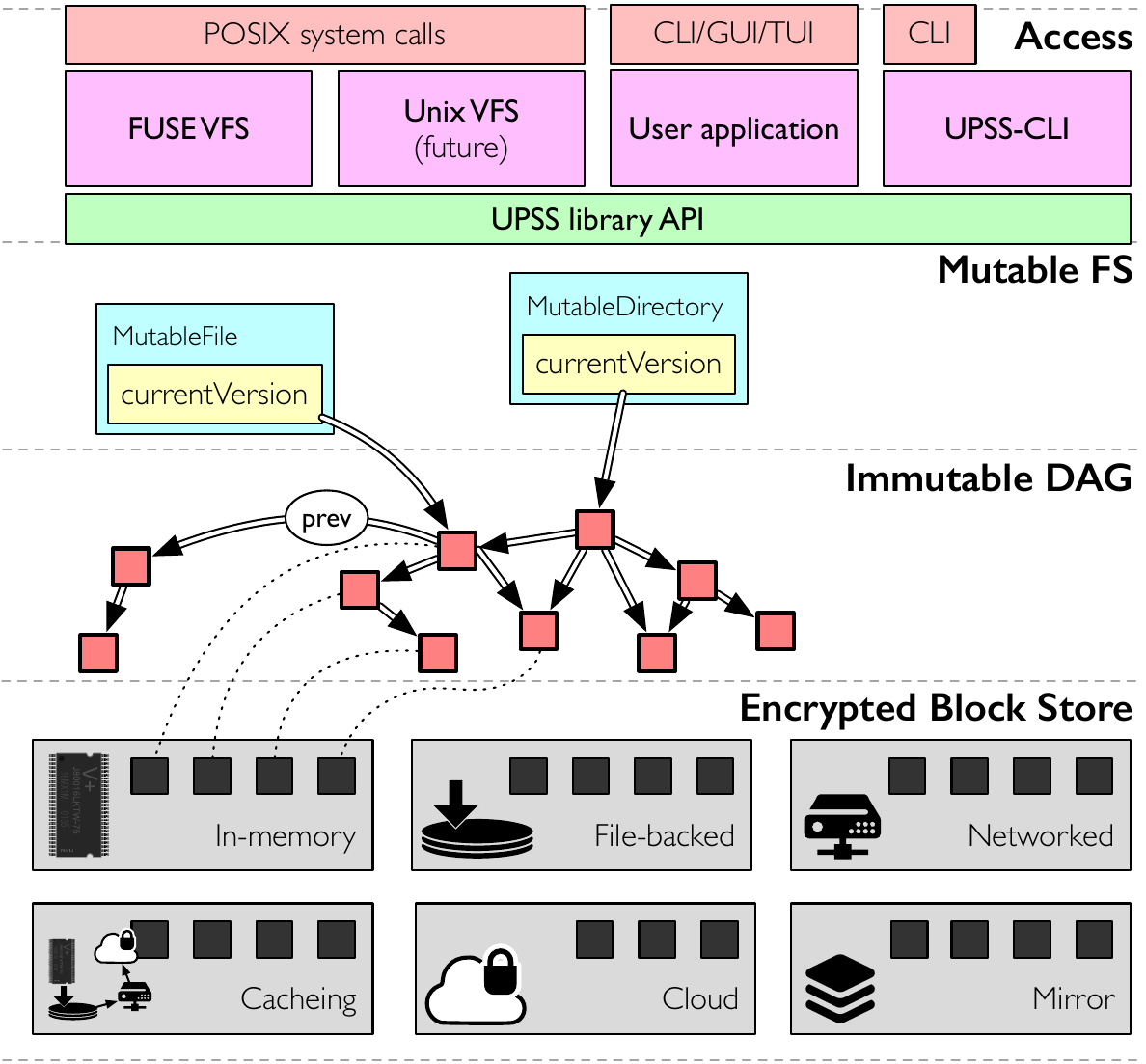}
	\caption[\systemname{} overview]{The layered structure of \systemname{}.}
	\label{fig:layers}
\end{figure}

In this section, we review key elements of the \systemname{} design, which was seeded in~\cite{upss,bozorgi2020online}
 and expaned in \cite{icissp24},  and
describe new design elements that enable practical performance
and novel applications that had previously been envisioned as future work.
These elements are visible across four layers
shown in \Cref{fig:layers}:
untrusted storage (\Cref{sec:blockstore}),
an immutable copy-on-write DAG of blocks (\Cref{sec:immutable}),
mutable filesystem objects (\Cref{sec:mutable}) and
two user-visible filesystem interfaces (\Cref{sec:file-access}).

\subsection{Untrusted storage}
\label{sec:blockstore}

Like all filesystems, \systemname{} ultimately stores data in fixed-size blocks
on persistent media.
Block sizes are all multiples of common physical sector sizes and are set by
the backing store rather than the client.
\systemname{} uses a default block size of 4\,kiB that can be overridden on a
per-store basis.
Unlike other filesystems, all \systemname{} blocks are encrypted
in transit and at rest: plaintext blocks is only held in memory and never
stored to persistent media. 
Rather than using per-file or per-filesystem encryption keys, each block is
encrypted with a key $k_B$ derived from its plaintext and
named by a cryptographic hash $n_B$ of its ciphertext. The 2-tuple $(n_B, k_B)$
constructs a \textit{block pointer} as given in \cref{eq:bp}. 

\begin{align}
\label{eq:bp}
\begin{split}	
k_B &= h(B) \\
n_B &= h \left( E_{k_B} \{ B \} \right) \\
\end{split}
\end{align}

In this equation, $B$ represents the plaintext contents of a block,
which contains user content and random padding to fill out the fixed-size block,
$h$ is a cryptographic hash function and $E$ is a symmetric-key encryption algorithm.
A block pointer is a cryptographic capability~\cite{capabilitysystems} to fetch,
decrypt and read a block's contents, though not to modify it, as blocks are
immutable.
Changing a single byte in the block would change a block's encryption key $k_B$,
which would change the encrypted version of the block,
which would change its name $n_B$.
As a matter of practical implementation, serialized block pointers also contain
metadata about their hashing and encryption algorithms
(typically SHA3~\cite{dworkin2015sha} and AES-128~\cite{nist:2001:aes}).

Deriving a symmetric encryption key from a block's contents is an example of
\textit{convergent encryption}~%
\cite{douceur2002reclaiming, li2013secure, agarwala2017dice}.
Convergent encryption is a symmetric-key encryption technique in which identical
ciphertexts are produced from identical plaintexts. 
This technique affords two benefits: a reduced burden of key management
and the possibility of block (rather than file) level data
deduplication~\cite{satyanarayanan1990coda,douceur2002reclaiming}.
Deduplication is an important feature for global-scale information sharing systems in which many users
may share the same content with others.
By deduplication, only two extra 4~KiB meta blocks are required to ingest a 1~GB file to
\systemname{} for the second time with the same content.
However, convergent encrypion and deduplication bring with themselves some risks that
are discussed in \Cref{sec:security-model}.

\subsubsection{Block stores}
A narrow API including \rust{read}, \rust{write}, \rust{block_size} and \rust{is_persistent}
methods is implemented by several types of block stores shown in
\Cref{fig:layers}:
in-memory (non-persistent), file-backed, networked,
cloud via Amazon S3 \cite{amazons3} or Azure blob storage \cite{azure},
caching and mirror. The caching and mirror block stores consist of multiple stores,
that accomplish different tasks. The former enables caching (\Cref{sec:caching})
and the latter handles replication (\Cref{sec:replication}), both at the block level. 

When an encrypted block is stored in a block store, the block store responds
with a block name $n_B$ derived using that store's preferred cryptographic hash algorithm.
A block's name can be used to retrieve the block in the future without any
further authorization
--- it is a cryptographic \textit{capability}~\cite{capabilitysystems}.
This approach allows block stores to be oblivious to user identities and content
ownership.
Instead, it is a \textit{content-addressed store}.
The operator of a block store cannot view plaintext content or even
directly view metadata such as file sizes or directory-file relationships.
Inference of these relationships is discussed in \Cref{sec:security-model},
which also describes the stronger privacy and security properties that
\systemname{} provides relative to other
cryptographic and conventional filesystems.

\subsubsection{Caching}
\label{sec:caching}
\begin{figure}
  \includegraphics[width=0.9\linewidth]{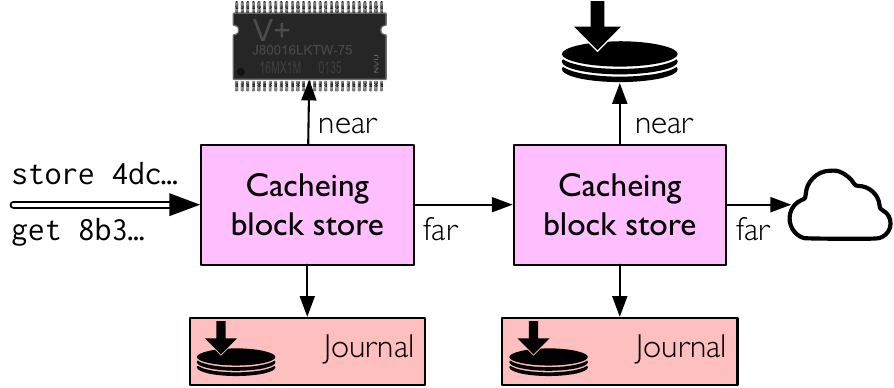}
  \caption{A caching hierarchy of untrusted block stores. \label{fig:cacheing}}
\end{figure}

The caching block store consists of two other near and far stores and a journaling
mechanism. A near store can be an in-memory block store that processes the operations
faster than a far store that can be a file-backed, networked, cloud, mirror, 
or another caching block store. Note that both near and far stores can be any block stores.
By having the caching block store, \systemname{} enables building a cache hierarchy
as shown in \Cref{fig:cacheing}.
For storing an encrypted block, the caching block store stores the block to the near
store and journal it to an on-disk file. The journaled blocks will be processed in
the background to be stored to the far store. For reading, the caching block store tries
to read the block from the near store and if it does not exist (e.g., the near store
is an in-memory store which has been cleared), the block is read from the far store.
The confidentiality and immutability of blocks in a block store enable
aggressive yet safe caching, even with remote storage on untrusted
systems. This makes \systemname{} achieve better performance results as discussed
in \Cref{sec:eval}.

A challenging problem with caching data in any information system is handling 
inconsistencies; a block's content can be updated in a cache while not in
other locations. However, \systemname{} avoids any cache inconsistencies
and reduces this problem to a version control problem by
the immutable nature and cryptographic naming of the stored blocks.
A block may be present within or absent from a store, but it
cannot be inconsistent between two stores: even the smallest
inconsistency in content would cause the blocks to have different
cryptographic names.

\subsubsection{Data availability via replication}
\label{sec:replication}
The mirror block store handles data replication across multiple block stores. 
For storing an encrypted block, the mirror block store replicates the block
to all block stores in parallel and returns the block name upon successful 
replication.
For reading, the mirror block store queries the block by its name from all
block stores in parallel and returns the block from a block store that 
responds faster and ignores other block store responses. 

\subsection{Immutable DAGs}
\label{sec:immutable}

\begin{figure}
  \centering
  \includegraphics[width=0.5\linewidth]{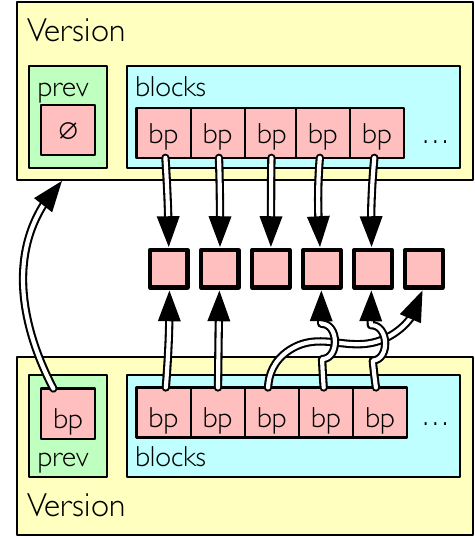}
  \caption{
    Two sequential \rust{Version} objects that reference
    two common blocks and one diverging block.
    \label{fig:versions}
  }
\end{figure}

\systemname{} uses directed acyclic graphs (DAGs) of immutable blocks to
represent files and directories.
Relationships among blocks are specified by \rust{Version}
objects that describe arbitrary-length collection of immutable blocks,
each accessible by their block pointers.
As shown in \Cref{fig:versions}, multiple \rust{Version} objects can
reference underlying immutable blocks, facilitating the copy-on-write
modification of files and directories described in \Cref{sec:mutable}.
\rust{Version} objects are themselves stored in \systemname{} blocks,
allowing them to be named according to \textit{their} cryptographic hashes.
For files smaller than 100 kiB, a \rust{Version} fits in a single \systemname{}
block.
A \rust{Version} may contain a block pointer to a previous
\rust{Version}, and thus a \rust{Version} can be used as a
Merkle tree~\cite{merkle:1979:thesis} (more precisely, a Merkle DAG)
that represents an arbitrary number of versions of an arbitrary quantity
of immutable content.

This use of Merkle DAGs reduces the problem of data consistency to that of
version control: it is possible for two files to
contain different blocks, but not two variations of the ``same'' block.
It is left to the user of these immutable DAGs to provide mutable
filesystem objects using copy-on-write (CoW) semantics and to ensure that new
blocks are appropriately pushed to backend block stores.

\subsection{Mutable filesystem objects}
\label{sec:mutable}

Conventional mutable filesystem objects (files and directories) are provided by 
\systemname{} by mapping arrays of bytes into mutable \rust{Blob} objects.
These objects maintain copy-on-write (CoW) references to underlying blocks
and versions.
Non-traditional objects such as structured binary key-value data structures
are also possible using multiple blobs and versions.

A \rust{Blob} manages an array of bytes via copy-on-write block references,
starting from an empty sequence of blocks and permitting operations
such as truncation, appending and random-access reading and writing.
A \rust{Blob} accumulates edits against an immutable \rust{Version} in an
``edit session''~\cite{Asklund:1999:extensional-versioning}
until a file or directory is persisted into a new immutable \rust{Version}.
This allows \systemname{} to accumulate write operations and batch them into
aggregate CoW operations.

\begin{figure}
	\centering
	\includegraphics[width=\columnwidth]{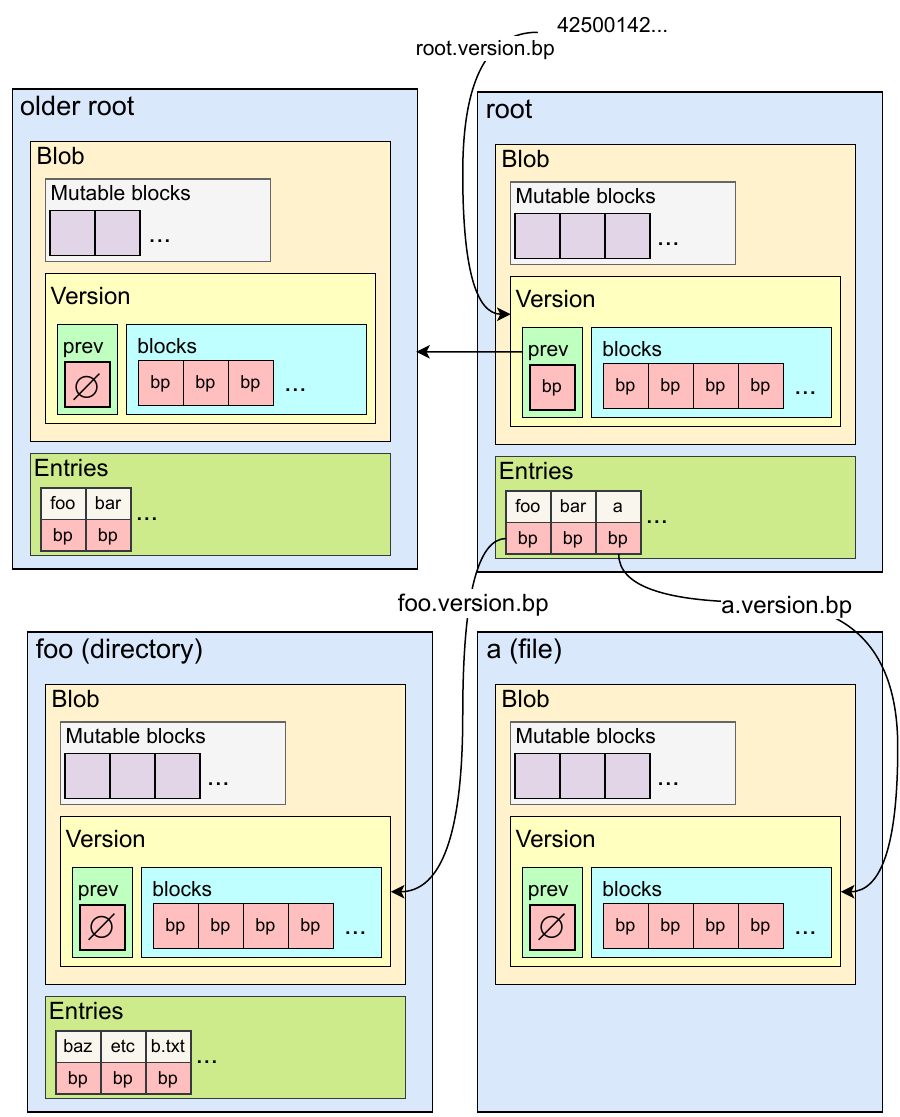}
	\caption[Directory hierarchy figure]{
    An example of a directory hierarchy in \systemname{}.
	}
	\label{fig:dir-hierarchy}
\end{figure}

Files and directories are both backed by \rust{Blob} objects, and both can be
explicitly persisted to backing storage via API calls
\rust{persist()} and \rust{name()}, which persists an object and
returns its block pointer.
A file version can be named by a block pointer to its \rust{Version} object
which represents the file's content and, optionally, history.
A directory is represented as a sequence of directory entries, each of which
maps a unique, user-meaningful name to a filesystem object (file or directory).
A directory can be persisted by serializing its entries into a \rust{Version}
that is named by a block pointer.
Thus, directories are also Merkle DAGs that reference the
lower-level Merkle DAGs of other file and directory objects.
\Cref{fig:dir-hierarchy} shows an example of a directory hierarchy in \systemname{}.
In this example, upon persisting, \textit{a}'s content is stored into encrypted
blocks and their block pointers are added to \textit{a}'s \rust{Version}
structure and the \rust{Version} is stored in encrypted blocks as well and its block pointer
is included in \textit{a}'s parent directory entries.

\begin{figure}
	\centering
	\includegraphics[width=0.8\columnwidth]{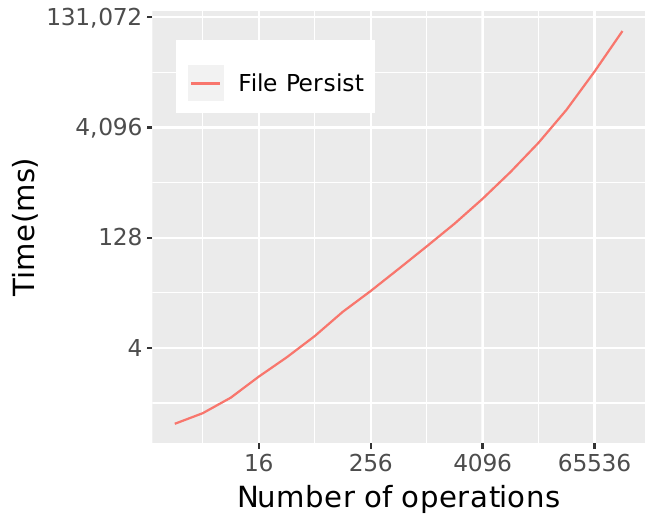}
	\caption[Persist time for different number of files]{
		The time required to persist files to a block store
    scales superlinearly with the number of files.
    Directory persist times are almost identical to those of files.
	}
	\label{fig:persist}
\end{figure}

Cryptographic hashes are computed and blocks encrypted when files and
directories are persisted,
making persisting one of the most expensive operations in \systemname{}.
Tracking chains of \rust{Version} objects in addition to content makes both the
time and storage requirements for persistence superlinear.
It is, therefore, only done when requested via the API
or, in the case of \fsname{}, every 5\,s.
The time required to persist 4\,kiB files after $n$ filesystem operations
is shown in \Cref{fig:persist}.
Based on our measurement results, the total space $s_t$ required in a block store to store $s$ bytes of content
follows \Cref{eqn:storage}.

\begin{equation}
s_t = \left( 1.09 + 0.001613\, s \right) s
\label{eqn:storage}
\end{equation}

\subsubsection{Mutation and versioning}
\label{sec:mutation-versioning}

Naming all filesystem objects by block pointers to \rust{Version} structures
introduce new challenges to handling modifications. Whenever a file or
directory is modified in a directory hierarchy, a new block pointer
is generated that should be updated in the object's parent entries
, and this update should be applied up to the root directory.
In order to handle updates efficiently, every file and directory object
keeps an \rust{Updater} object, which is a reference to its parent in-memory object.
Upon modification and persisting, an object notifies its \rust{Updater} about
its new block pointer and the parent object is modified to reflect the child's
new version.
Similar requirements for updating of parents exist in other CoW filesystems
such as ZFS~\cite{bonwick2003zettabyte}, but the case of a global CoW filesystem
such as UPSS is more challenging than that of local filesystems.
In a local CoW filesystem, it is possible for the filesystem implementation to
be aware of all concurrent uses of a parent directory, including by multiple
users.
In a global filesystem, however, not all uses of a parent directory are visible
to a local host.
UPSS therefore, treats every update to a filesystem subtree as a potential
versioning operation, allowing new directory snapshots to be created and
shared as described in \cref{sec:snapshot}; versions can be integrated at the
level of filesystem interfaces as described in \cref{sec:file-access}.

\subsubsection{Snapshot and sharing}
\label{sec:snapshot}
As a copy-on-write filesystem, \systemname{} provides cheap snapshots of
previous versions.
\systemname{} creates snapshots whenever requested via \ccode{sync(2)},
\ccode{fsync(2)} or the UNIX \command{sync(1)} command, or in case of FUSE 
(filesystem in user space) wrapper, by querying
a directory's cryptographic name with POSIX extended attributes, or in case of \systemname{}
API, by calling \rust{persist()} or \rust{name()} methods. 
As shown in \Cref{lst:xattr-hash}, extended attributes can be queried to
retrieve the cryptographic hash or serialized block pointer of any
\systemname{} file or directory.
Exposing serialized block pointers to users facilitates sharing of file
and directory snapshots from user to user, including sharing from
\systemname{} FUSE wrapper snapshots to users employing the \systemname{} CLI.
Also, this allows users to check integrity guarantees over
file and directory Merkle DAGs, facilitating blockchain-like applications.

As a user-empowering \textit{sharing} system, these snapshots can be quickly
shared with other users for read-only access: user $a$ need only share the block
pointer to a file or directory with user $b$, and user $b$ will be able to
retrieve the content from a block store and decrypt it.
Since block pointers correspond to immutable blocks, user $b$ cannot modify the shared block.
Upon modification, a new block is generated with a new block pointer and
user $a$ still has access to the unmodified shared block.

\begin{figure}[htb]
\begin{lstlisting}[label=lst:xattr-hash, caption={Retrieving cryptographic file
information via a POSIX extended attribute and use it in the UPSS CLI.}, captionpos=b]
% attr -g hash mnt/some-dir
sha3-512:hdd3P80hjERoF1PO9ezuOEQQwG/Goey2Up5je...
% attr -g bp mnt/some-file
4250014222000000000000018e01011e47605fef888cc6...
% upss --store=store.dat get 42500142220000000...
This is some file content!
\end{lstlisting}
\end{figure}

\subsection{File access interfaces}
\label{sec:file-access}

Users can access an \systemname{} filesystem via a variety of interfaces,
including a Rust API which can be compiled to WebAssembly,
a command-line interface and
a FUSE (Filesystems in Userspace) interface.
Unlike many filesystems, any \systemname{} directory can be treated
as the root directory of a filesystem.
Within a directory hierarchy, a user may persist any subdirectory to retrieve
a block pointer to an immutable version of it.
That version may then be used as the basis for further filesystem operations
including mounting, mutating and further sharing.
When new versions of files and directories are generated, parent directories
are updated until a new root directory version is created.
Storing the block pointer of that new root directory is the responsibility
of an \systemname{} client (API, CLI, FUSE or future native VFS implementation).
The \systemname{} CLI and FUSE clients both store 
this information in a local passphrase-protected file as per
PKCS \#5 Version 2.0 \cite{rfc2898} for interoperability.
\Cref{tbl:cli-commands} shows a list of commands that are supported by \systemname{}
CLI. 

\begin{table}
	\caption{A list of commands available via \systemname{} CLI.}
	\label{tbl:cli-commands}
	\centering
	\begin{tabular}{p{0.27\columnwidth}|p{0.6\columnwidth}}
    \toprule

			\multicolumn{1}{c|}{\textbf{Command}} & 
			\multicolumn{1}{c}{\textbf{Description}} \\
			\midrule

			\rust{upss init} & Initialize an empty filesystem \\
			\rust{upss ls} & List the files at a particular path \\
			\rust{upss info} & Verbose information about a path \\
			\rust{upss touch} & Create a file at a path \\
			\rust{upss mkdir} & Create a directory at a path \\
			\rust{upss append} &  Append to a file\\
			\rust{upss store} & Store a file at a path within UPSS \\
			\rust{upss history} & Prints a history of the file revisions \\
			\rust{upss name} & Get a file's block pointer in a path \\
			\rust{upss names} & List the file block pointers in a path \\
			\rust{upss get} & Get an UPSS file's content \\
			\rust{upss get-path} & Get file's name by its block pointer \\
			
			\hline
		\end{tabular}
\end{table}

\begin{figure}
	\centering
	\includegraphics[width=1\columnwidth]{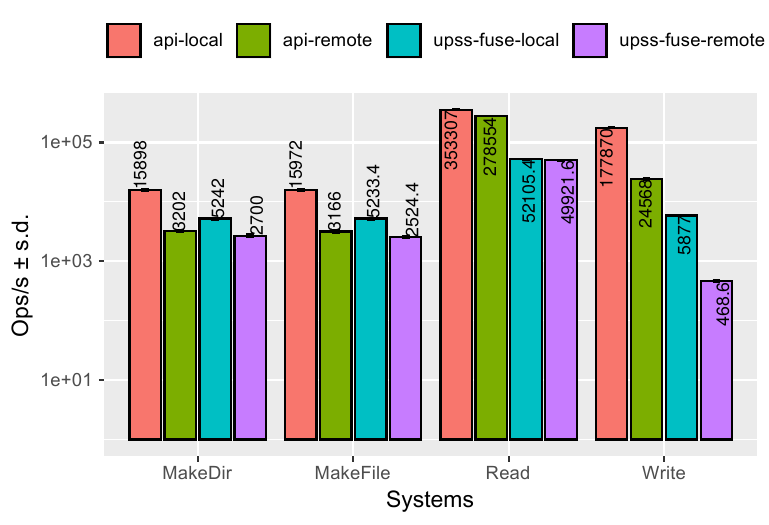}
	\caption[\systemname{} performance as API, local and remote filesystem]{
    \systemname{} performance when accessed via its API and via \fsname{}
    connected to a local or remote block store.
    The average number of operations per 60 seconds is reported for five runs;
    error bars show standard deviation.
	}
	\label{fig:api-local-remote}
\end{figure}

Direct API invocation provides clear performance benefits when compared to
FUSE-based wrapping.
As shown in \Cref{fig:api-local-remote}, directly invoking the
\systemname{} API yields higher performance than using a FUSE wrapper with
the same storage backend.
For two of the four microbenchmarks described in \Cref{sec:bench-desc},
the cost of the FUSE wrapper exceeds
that of the cost of communicating with a remote blockstore via direct \systemname{} API.

\section{Security model}
\label{sec:security-model}
\systemname{} is designed to provide a new approach to private data storage
and sharing, enabling both strong security properties and simple sharing
across systems and users.

Other cryptographic filesystems take a variety of approaches to encryption.
EncFS~\cite{encfs} and NCryptFS~\cite{wright2003ncryptfs} employ encryption
for file content but not the filesystem itself, e.g., directory structure.
Systems such as CryFS~\cite{messmer2017novel} protect the entire filesystem
with a single encryption key, which has two implications.
Firstly, it implies strong filesystem boundaries and precludes safe subset
sharing: the unit of possible sharing in the system is a filesystem, not a file.
This may not match a user's desired sharing granularity, in which files or
directory-oriented bundles of content may be passed among multiple users or
systems.
Secondly, this coarse-grained use of encryption increases the value of one
specific encryption key, making it a more attractive target for attackers.
In contrast to these systems, \systemname{} encrypts data using per-block keys
derived from block content.
This removes the need for separate storage of keys as security metadata and
reduces the value to an attacker of any single encryption key.
Encryption keys in \systemname{} are not considered secret to authorized users:
a user who is authorized to read a file is authorized to learn the key that was
used in the reading process since that key is not used to protect other
information.

Commonly-used cloud-based storage systems allow storage providers to examine
users' plaintext directly.
In contrast to these systems, backend block store providers in
\systemname{} are only able to see a sea of encrypted blocks from users.
Encryption is performed on the client side, so users' software can ensure that
plaintext is never exposed to storage providers.
In addition to hiding file content, this sea-of-encrypted-blocks approach to
backend storage ensures that metadata such as file sizes and directory
structures are not revealed explicitly to storage providers.
Providers could perform traffic analysis to infer relationships among various
blocks, but only at significant computational cost.
Even this threat could be addressed by the use of oblivious transfer techniques
such as ORAM, but other large-scale distributed systems that support such
techniques have disabled them due to unwarranted cost~\cite{utahblospost}.

In addition to inferring relationships among encrypted blocks, it is also
possible for a malicious block store to attack the availability of blocks by
refusing to serve them when requested, or by delaying that service, or by
serving incorrect blocks, however, the client can independently
verify their correctness by checking their cryptographic hash against the
received blocks.
Therefore, \systemname{} does not rely on a consensus of storage nodes.

Convergent encryption, first used in Farsite~\cite{adya2002farsite},
provides clear benefits to \systemname{} in terms of de-duplication and
key management, but it can also introduce risks that are not present
in traditional cryptosystems.
Convergent encryption is a deterministic encryption model, but the traditional
objective of indistinguishability under chosen plaintext attack (IND-CPA)
forbids determinism in encryption.
Convergent encryption can therefore be used to reveal
whether or not a given plaintext has previously been encrypted and stored
in the content addressable storage:
an attacker can encrypt a plaintext, present the ciphertext to a block
store and use timing or other response information to determine whether that
block has previously been stored.
Worse still, naïve forms of convergent encryption would allow an attacker to
guess variations on a known format (\data{user=1000}, \data{user=1001}, etc.)
to test whether any such variations have previously been stored.

\systemname{} addresses these concerns by appending random padding to
plaintext blocks to bring them to the fixed block size.
Small blocks of user data, those that most need protection from guessing
attacks, are padded with high-entropy random bits.
Full blocks of user data, such as content from shared media files,
do not require padding, allowing such files to enjoy deduplication.
Confirmation of a large existing block is still possible, but not via guessing
attacks due to the block sizes involved. Also, identification of the users
who have stored a particular block is protected: no user can be associated with the content.
By default, any block of plaintext data that is smaller than the fixed block
size will have random padding appended, although it is also
possible to employ fully deterministic encryption if the known weakness of
convergent encryption are not a concern.
For example, a typical AWS credential file with a known access key will have
$s_K = 72$\,B of data that could be known to the attacker and
$s_s = 40$\,B of Base64-encoded secret key that the attacker
would like to guess.
\systemname{} will append $s_p = 3,984$\,B of random padding to this data
on encryption, which is almost 32,000 high-entropy bits.
A brute-force attack against such a block, containing a known access key,
may be attempted using a fixed $s_K = 72$\,B and varying the other 4,024\,B.
This attack would be expected to succeed after $2^{32,191}$ attempts,
as shown in \cref{eq:guesses}.
This compares favourably to the $2^{239}$ attempts that would be required to
brute-force the AWS secret key itself.

\begin{equation}
\label{eq:guesses}
\begin{split}
\mathbb{E}_\textrm{guesses}
& = \frac{1}{2} \cdot 2^{ 8 \left( s_s + s_p \right) } \\
& = 2^{ 8 \left( 40 + 3,984 \right) - 1 } \\
& = 2^{32,191}
\end{split}
\end{equation}

\systemname{} does not explicitly represent users or user identities.
This allows applications or clients to bring their own user model to the
filesystem and avoid the limitations of system-local users, as in systems
that project local filesystems to a multi-system context,
e.g., multi-user EncFS~\cite{leibenger2016encfs}).
The \systemname{} FUSE interface allows a single user to mount an
\systemname{} directory and manipulate it like an external drive; the
\systemname{} CLI allows any user with local permissions for the file
containing the root block pointer to update it accordingly.
Multiple users on separate systems can share a backend block store without
interference, but the common block store permits efficient sharing among
systems and users.
Higher-level applications can bring their own concepts of users and sharing
semantics to the \systemname{} filesystem: application channels can be used
to share block pointers to content and new versions of content can be shared
back.

The judicious use of full-filesystem, per-block convergent encryption allows
\systemname{} to employ untrusted storage backends that can scale to the
largest of workloads without revealing user data or metadata.
Its user-agnostic approach allows it to be employed within applications and in
a range of uses from a local filesystem to a global sharing system.

\section{Performance evaluation}
\label{sec:eval}
In this section, we demonstrate the practicality of \systemname{} as a
local filesystem (\Cref{sec:local-fuse}),
a network filesystem (\Cref{sec:network-fuse})
and a global filesystem (\Cref{sec:global-fuse}).
Although \systemname{} achieves its best performance when accessed via API
rather than FUSE, employing the FUSE interface allows us to directly compare
its performance with the performance of extant systems.
These performance comparisons are completed using a suite of microbenchmarks
and one FileBench-inspired macrobenchmark.

\subsection{Benchmark description}
\label{sec:bench-desc}

We have compared the performance of \systemname{} with other systems using both
custom microbenchmarks and a Filebench-inspired benchmark.
All benchmarks were executed on a 4-core, 8-thread 3.6~GHz Intel Core-i7-4790
processor with 24~GiB of RAM and 1~TB of ATA 7200 RPM magnetic disk, running
Ubuntu Linux 4.15.0-72-generic.
Remote block stores, where employed, used machines with different
configurations as described in \Cref{sec:network-fuse}.

For microbenchmarking, we evaluated the cost of
creating files and directories and reading and writing from/into
on-disk local and remote block stores.
For evaluating file and directory creation, we generated a user-defined number
of files and directories, added them to an ephemeral root directory and
persisted the results into file-backed block stores.
To evaluate read and write operations, we generated 1000 files filled with
random data of size 4~KiB, the natural block size of our underlying storage,
select a file randomly and performed sequential read and write operations on it.

We also implemented a macrobenchmark that simulates more complex behaviour.
In this benchmark, we selected a file randomly from a set of files and
performed 10 consecutive read and write operations with different I/O sizes:
4~KiB, 256~KiB, 512~KiB and 1~MiB.
The building blocks of this benchmark were inspired by the
Filebench framework~\cite{filebench}, but Filebench itself could not produce
the fine-grained timing information used to produce the figures shown in
\Cref{sec:local-fs,sec:network-fuse,sec:global-fs-eval}.

\subsection{\systemname{} as a local filesystem}
\label{sec:local-fuse}
\begin{figure}[ht]
	\centering
	\includegraphics[width=1\columnwidth]{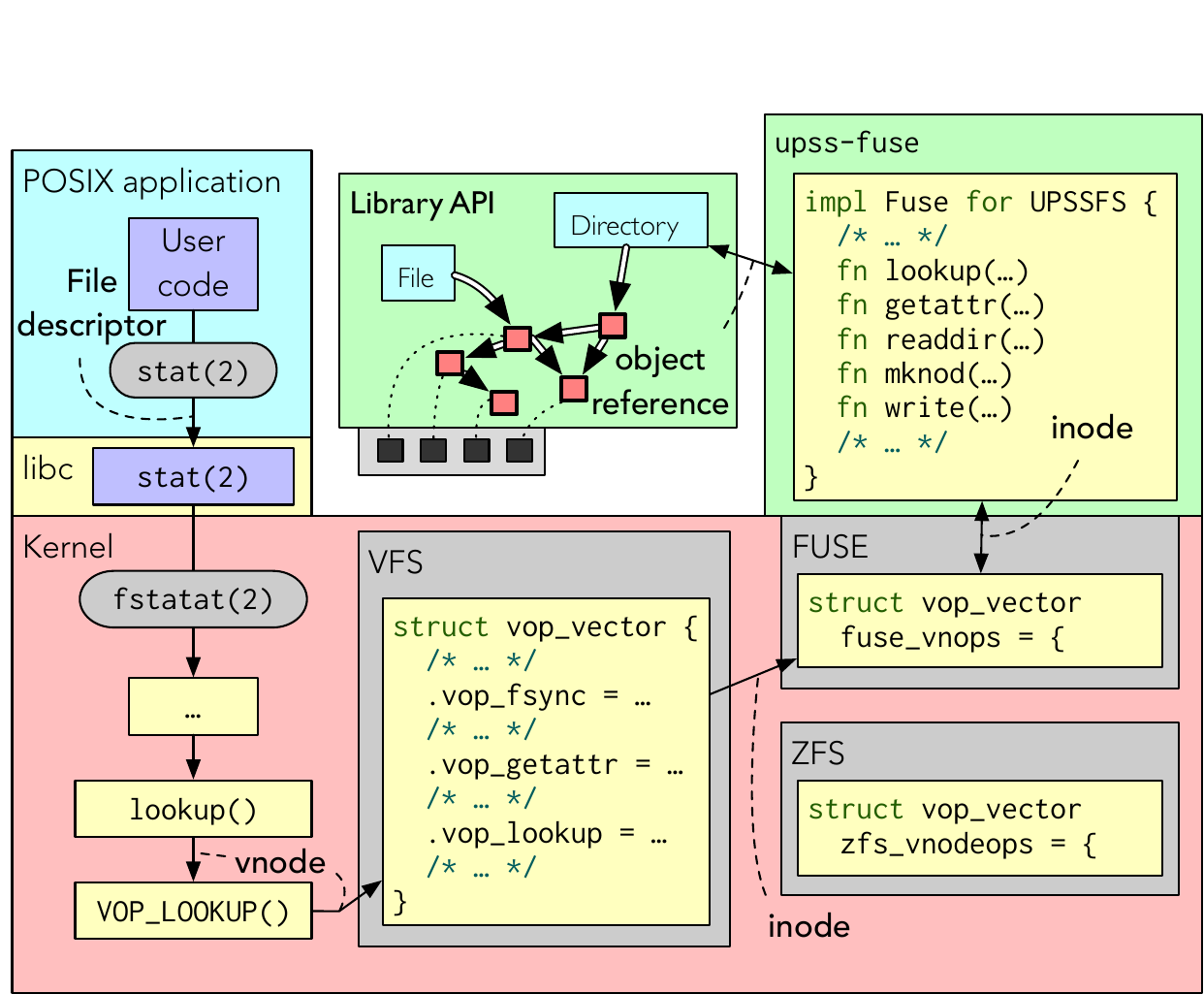}
	\caption[FUSE and \fsname mapping]{
		\fsname{} exposes a \systemname{} directory to POSIX applications via
		an in-kernel FUSE device.
	}
	\label{fig:fuse}
\end{figure}

Direct usage of the \systemname{} API requires program modification --- and,
today, the use of a specific programming language.
In order to expose the benefits of \systemname{} to a wider range of software,
we have implemented a \textit{filesystem in userspace (FUSE)}~\cite{fuse}
wrapper that exposes \systemname{} objects to other applications via a hook
into the Unix VFS layer. The challenge here is picturing \systemname{}'s
global view of encrypted blocks to a local view of files and directories
that can be accessed via FUSE inode numbers. To tackle this, \fsname{} uses 
a mapping from FUSE inode numbers to
in-memory \systemname{} objects to service VFS requests, as shown in \Cref{fig:fuse}.
This allows conventional applications to access an \systemname{} directory
mounted as a Unix directory with POSIX semantics, though there is one
unsupportable feature: hard links.
Hard links are defined within the context of a single filesystem, but
\systemname{} is designed to allow any directory to be shared as a root
directory of a filesystem.
Owing to this design choice, it is impossible to provide typical hard link
semantics and, e.g., update all parents of a modified file so that they can
perform their own copy-on-write updates (see \Cref{sec:mutable}).
Therefore, we do not provide support for hard links --- a common design choice
in network file systems such as NFS.

The \fsname{} wrapper exposes an ephemeral plaintext view of an
\systemname{}'s directory underneath a Unix mount point, allowing conventional
file and directory access, while keeping all data and metadata encrypted at
rest in a local or remote block store (see \Cref{sec:blockstore}).
Unlike existing cryptographic filesystems such as
NCryptFS~\cite{wright2003ncryptfs} and EncFS~\cite{leibenger2016encfs,encfs},
no plaintext directory structure is left behind in the mount point after the
filesystem has been unmounted.

\subsubsection{Consistency}
\label{consistency}
\begin{figure}
	\centering
	\includegraphics[width=0.8\columnwidth]{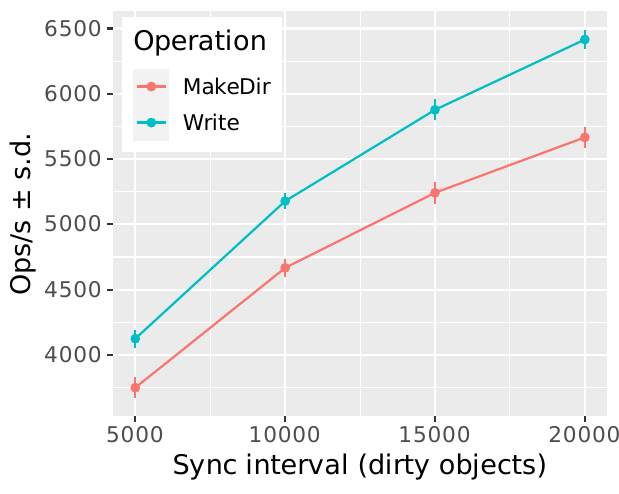}
	\caption[Trade-off of sync frequency vs performance]{		
    Performance of 4\,kiB operations vs sync frequency (in number of dirty
    objects) over five runs.
	}
	\label{fig:persist-tradeoff}
\end{figure}

In order to provide data consistency, \fsname{} requests that \systemname{}
persist a ``dirty'' --- i.e., modified --- root directory every five seconds,
or after a tunable number of dirty objects require persisting.
As described in \Cref{sec:mutable}, persisting a \object{Directory} object
causes its versioned children to be recursively persisted (if dirty), after which the cryptographic block pointer for the new root directory version can be stored in the
\fsname{} metadata file. This root block pointer is the only metadata that \fsname{} needs to mount the filesystem again. The block pointer size is 80~bytes as the defualt hashing and encryption algorithm in \systemname{} are SHA3-512 and AES-128 respectively. 
As in other copy-on-write filesystems, the cost of persisting an entire
filesystem depends on the amount of ``dirty'' content in the filesystem.
The trade-off between the demand for frequent data synchronization and the
requirement for more frequent --- though smaller --- persistence operations
is illustrated in \Cref{fig:persist-tradeoff}.

\subsubsection{Performance comparisons}
\label{sec:local-fs}
To illustrate the performance of \systemname{} when used as a conventional
local filesystem, we compared \fsname{} with the cryptographic
filesystems CryFS~\cite{messmer2017novel} and
EncFS~\cite{leibenger2016encfs, encfs}, also based on FUSE, 
as well as the mature, heavily-optimized ZFS~\cite{bonwick2003zettabyte}.
ZFS is not a cryptographic
filesystem designed for fine-grained confidentiality,
but it does share some design elements with \systemname{}:
it is a log-structured filesystem with copy-on-write updates
that uses cryptographic hashes to name blocks.
In contrast to \fsname{}, ZFS has been extensively optimized over the
past two decades to become a high-performance, widely-deployed filesystem.

We mounted each of these four filesystems on different paths in the Linux host
referenced in \Cref{sec:bench-desc} and ran four microbenchmarks to test
their speed in creating empty directories (\textbf{MakeDir}),
creating empty files (\textbf{MakeFile}),
reading randomly select files sequentially including 4~KiB of data (\textbf{ReadFile})
and writing random data to files (\textbf{WriteFile}).

\begin{figure}
	\centering
	\begin{subfigure}{.5\columnwidth}
		\centering
		\fbox{\includegraphics[width=0.9\columnwidth]{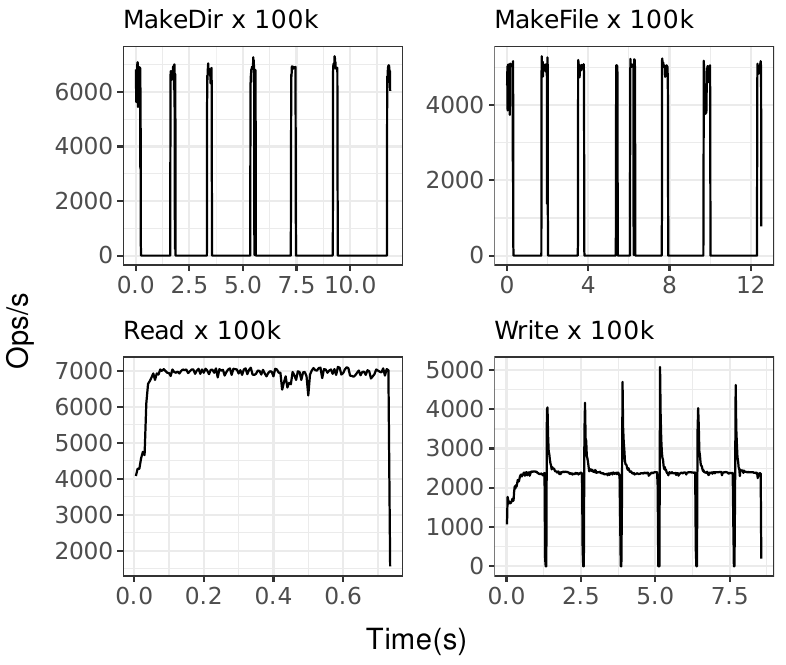}}
		\caption{\fsname{}}
		\label{fig:logtime-upss}
	\end{subfigure}%
	\begin{subfigure}{.5\columnwidth}
		\centering
		\fbox{\includegraphics[width=0.9\columnwidth]{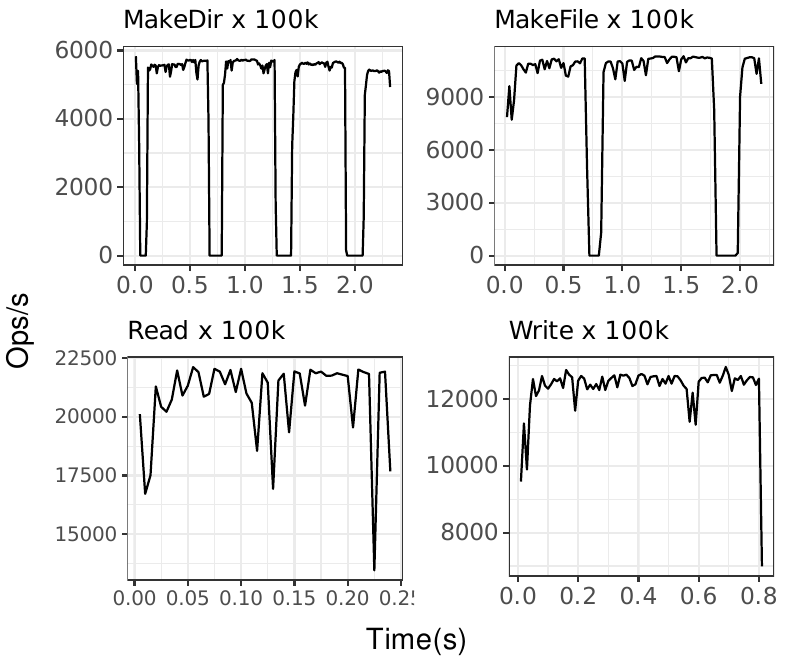}}
		\caption{ZFS}
		\label{fig:logtime-zfs}
	\end{subfigure}
	\\
	\begin{subfigure}{.5\columnwidth}
		\centering
		\fbox{\includegraphics[width=0.9\columnwidth]{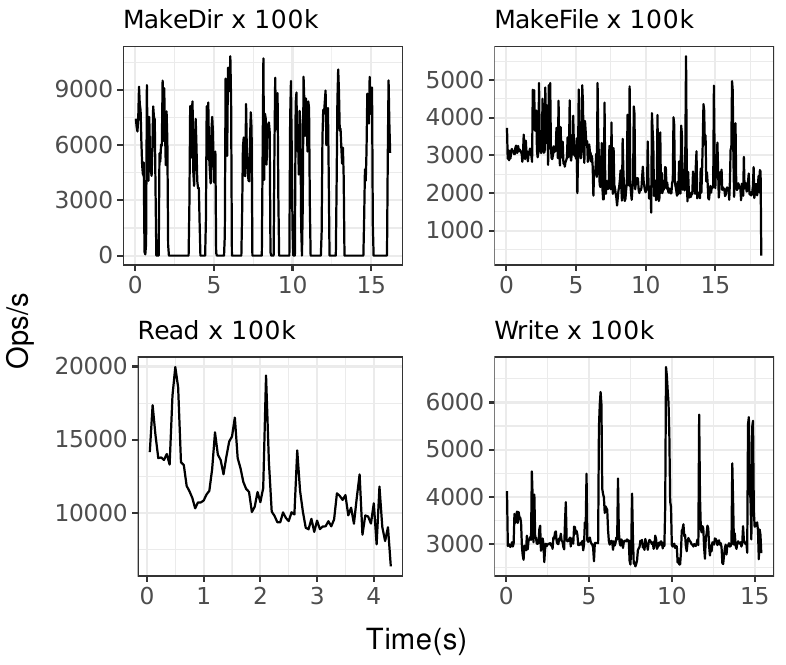}}
		\caption{EncFS}
		\label{fig:logtime-encfs}
	\end{subfigure}%
	\begin{subfigure}{.5\columnwidth}
		\centering
		\fbox{\includegraphics[width=0.9\columnwidth]{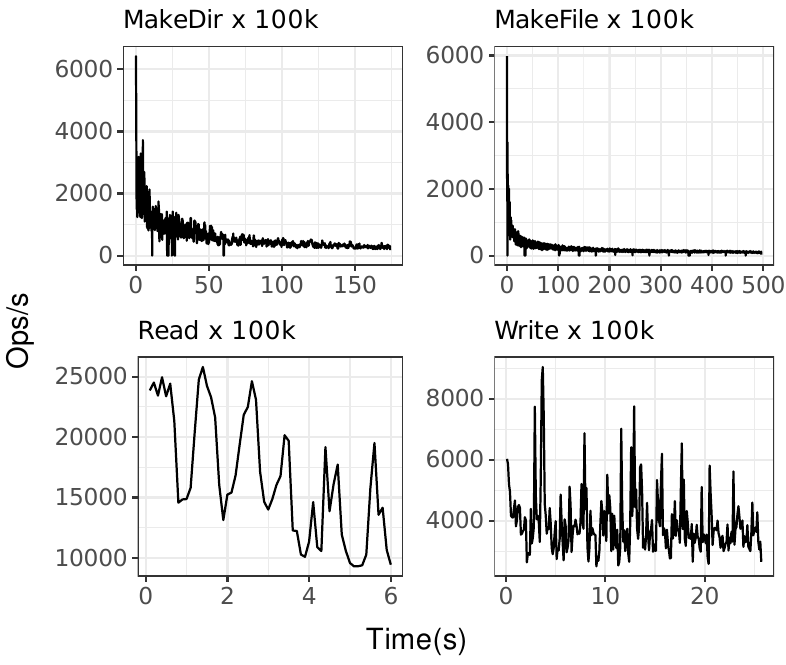}}
		\caption{CryFS}
		\label{fig:logtime-cryfs}
	\end{subfigure}%
	\caption[The local filesystems' behaviour]{
    Performance comparison of \fsname{} with CryFS, EncFS and ZFS.
    Benchmarks were run for 100\,kops.
	}
	\label{fig:logtime}
\end{figure}

Each of these four operations was run 100k and the behaviour of the
filesystems were reported in \Cref{fig:logtime}. In these plots, the $x$-axis represents the time needed to complete all 100k operations.
\systemname{} outperforms EncFS and CryFS for all operations, with performance
especially exceeding these existing systems in the
critical read and write benchmarks.
As might be expected, ZFS significantly outperforms \systemname{} in all benchmarks, with read performance $3 \times$ and
write performance $10.9 \times$ faster than \fsname{}.
In \fsname{}, creating files and directories have the same cost, as they are
both backed by empty collections of blocks.
We also note that \fsname{} performs $1.47-41.6 \times$ more operations per
second in various benchmarks than CryFS and EncFS while also providing
stronger security properties (see \Cref{sec:ralated-work}). This is due to our design choice that the requests are served from the mapped in-memory objects that are persisted periodically, if dirty. Therefore, expensive persist operations can be done quickly: with little
accumulation of dirty state, less synchronous persistence work is required.

These plots show the bursty nature of real filesystems, and in the case of
CryFS, they reveal performance that scales poorly as the number of requested
operations increases. Much of the bursty nature of these plots derives from how each
filesystem synchronizes data to disk.
For example, by default, ZFS synchronizes data every 5\,s or when 64\,MiB of
data has accumulated to sync, whichever comes first.
Similarly, to provide a fair comparison, \fsname{} is configured to synchronize after 5\,s or 15,000 writes (close to 64\,MiB
of data when using 4~KiB blocks).
These periodic synchronizations cause performance to drop, even on dedicated
computers with quiescent networks and limited process trees.

\subsubsection{Macro-benchmark}
\label{sec:macro}
\begin{figure}[ht]
	\centering
	\includegraphics[width=0.9\columnwidth]{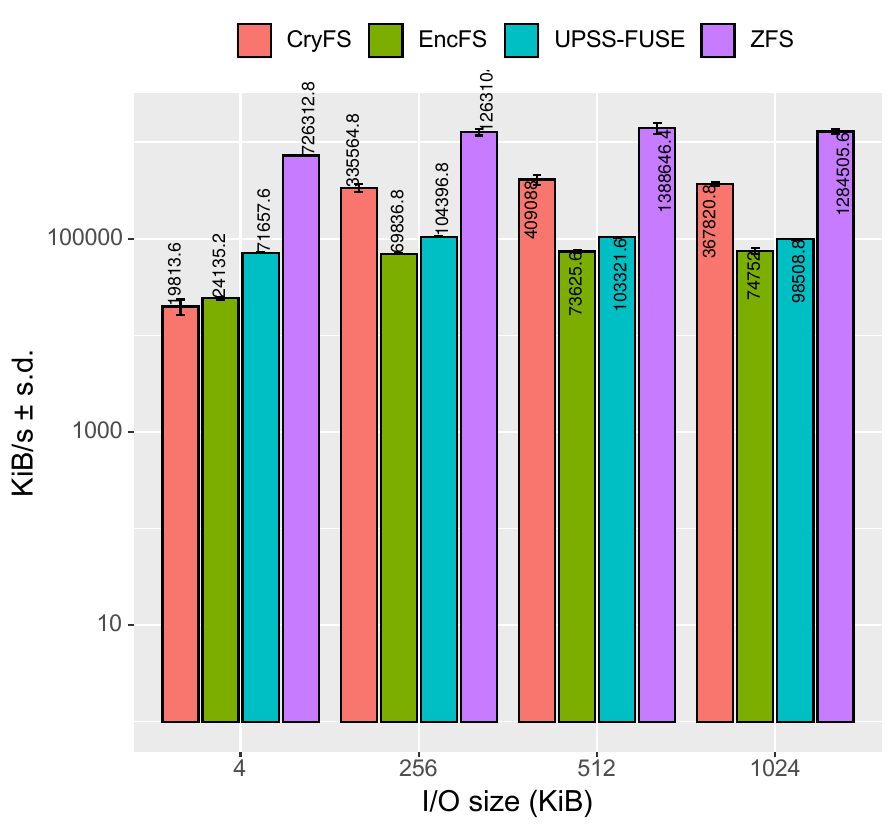}
	\caption[Local filesystems' performance for macrobenchmarks]{
		Performance of CryFS, EncFS, \fsname{} and ZFS for the macrobenchmark. The numbers are the average of KiB of I/O per second for
		five runs, each 60 seconds along with their standard deviations.
	}
	\label{fig:macro}
\end{figure}

We ran the macrobenchmark described in \Cref{sec:bench-desc} on \fsname{}, CryFS, EncFs and ZFS, to evaluate \fsname{} in a simulation in which consecutive read and write operations with different I/O sizes are performed on different files.
The results are reported in \Cref{fig:macro}.
As in our microbenchmarks, ZFS outperforms the other filesystems for different I/O sizes.
\fsname{} achieved better results than CryFS and EncFS for the 4~KiB case.
However, as the I/O size increases, CryFS outperforms \fsname{}.
The larger the I/O operation, the more fixed-sized blocks are generated by \fsname{}, each of which needs to be encrypted with a different key and persisted.
In CryFS, however, all the fixed-size blocks related to a file are encrypted with the same symmetric key.
This causes better performance for larger files, but at the same time makes CryFS inapplicable to the partial sharing and redaction use cases that can be supported by \systemname{}.
\systemname{} has been designed for small block sizes (typically 4\,kiB), as
decades of research has shown that filesystems mostly contain small files~%
\cite{rosenblum1992design,baker1991measurements,lazowska1986file}.

\subsection{\systemname{} as a network filesystem}
\label{sec:network-fuse}

Although \systemname{} can be used as a local filesystem, it is primarily
designed as a system for sharing data across networks with untrusted
storage providers.
\systemname{}' use of encrypted block stores, in which
confidentiality and integrity of these blocks' content are assured by
clients and not servers, allows us to build a block store in which a centralized
server exploits high-quality network links to transfer large numbers of
encrypted blocks --- the data plane --- regardless of what block
pointers are shared between users --- the control plane.
This design is amenable to multi-layer caching, as described in
\Cref{sec:blockstore}.
Thus, we have compared the performance of \fsname{} when connected to a
remote block store to that of SSHFS \cite{sshfs} and the venerable NFS \cite{nfs}.

\subsubsection{Performance comparison}
\label{sec:network-fs-eval}
As in \Cref{sec:local-fs}, we evaluated the performance of \systemname{} by
mounting an \fsname{} filesystem in a Unix mount point and
comparing it to other filesystems using four microbenchmarks.
In this section, however, we connected our \fsname{} filesystem to a remote
block store and compared our performance results against two other remote
filesystems: the FUSE-based SSHFS \cite{sshfs} and the venerable NFS \cite{nfs}.
Similar to \Cref{sec:local-fs}, one comparison filesystem is primarily designed for security
and the other has higher performance after a long history of
performance optimization.

\begin{figure}
	\centering
		\begin{subfigure}{.5\columnwidth}
			\centering
			\fbox{\includegraphics[width=0.9\columnwidth]{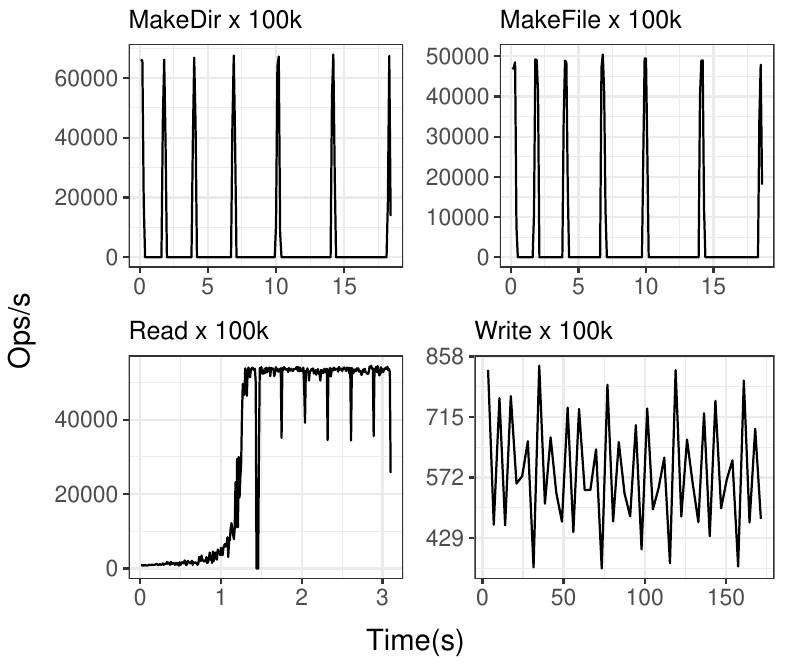}}
		\caption{\fsname{}-network}
		\label{fig:logtime-upss-network}
	\end{subfigure}%
	\begin{subfigure}{.5\columnwidth}
		\centering
		\fbox{\includegraphics[width=0.9\columnwidth]{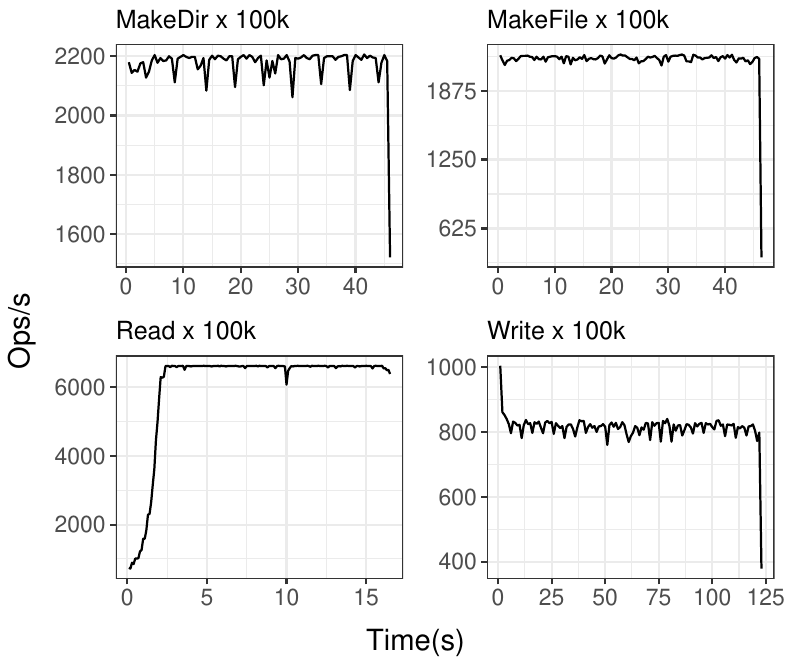}}
		\caption{NFS}
		\label{fig:logtime-nfs}
	\end{subfigure}%
	\\
	\begin{subfigure}{.5\columnwidth}
		\centering
		\fbox{\includegraphics[width=0.9\columnwidth]{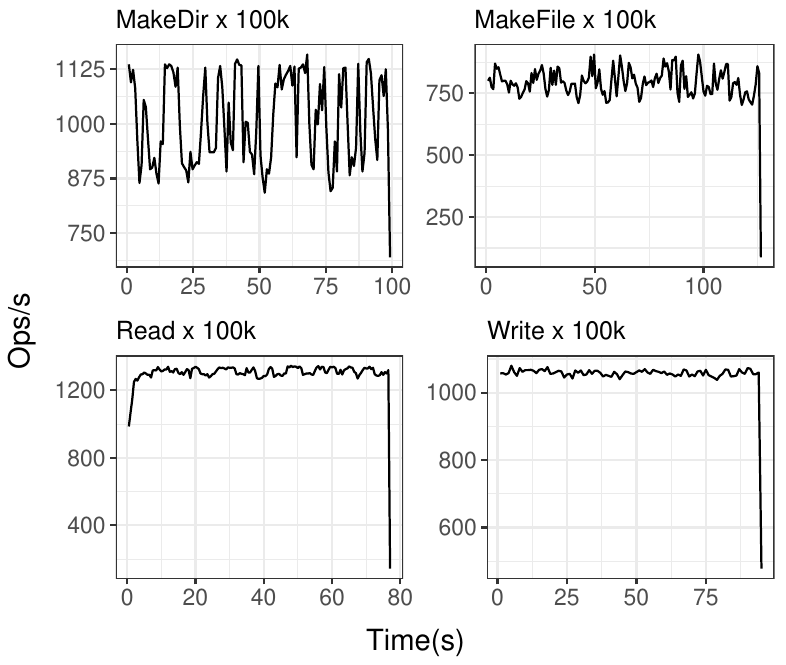}}
		\caption{SSHFS}
		\label{fig:logtime-sshfs}
	\end{subfigure}
	\caption[Network filesystems' behaviour]{
    Performance comparison of \fsname{}-network, NFS and SSHFS.
    Benchmarks were run for 100\,kops.
	  \label{fig:logtime-network}
  }
\end{figure}

The remote block store server was run on a 4-core, 2.2~GHz Xeon E5-2407 processor with 16~GiB of RAM and 1~TB of magnetic disk, running FreeBSD 12.1-RELEASE. The client machine, that runs \fsname{}, is a 4-core, 3.5~GHz Xeon E3-1240 v5 processor with 32~GiB of RAM and 1~TB of magnetic disk, running Ubuntu Linux 16.04.
The client and server were connected via a dedicated gigabit switch.
\Cref{fig:logtime-network} shows the behaviour of the benchmarked filesystems
when executing 100k \textbf{MakeDir}, \textbf{MakeFile}, \textbf{Read}
and \textbf{Write} operations.

\systemname{} outperforms SSHFS and even NFS for \textbf{MakeDir},
\textbf{MakeFile} and \textbf{Read} operations and for \textbf{Write}, it
achieves comparable results.
For the \textbf{Read} benchmark, \fsname{} has a slow start as encrypted blocks
are read from the remote block store and are loaded into memory.
After files are loaded into memory, other read operations are served from the
in-memory objects.
This causes \fsname{} to be about $5 \times$ faster than NFS in the
\textbf{Read} benchmark, validating \fsname{}'s approach to encrypted block
storage and the safe and aggressive caching it enables.

\subsection{\systemname{} as a global filesystem}
\label{sec:global-fuse}
In addition to local and network filesystem, \fsname{} can also be connected to untrusted cloud storage providers. To do so, we have implemented an \systemname{} block store backed in the Amazon S3 service \cite{amazons3} and compared its performance with S3FS \cite{s3fs}, Perkeep \cite{perkeep} and UtahFS \cite{utahfs,utahblospost}.

\subsubsection{Performance comparison}
\label{sec:global-fs-eval}
We mounted \fsname{} backed with the Amazon block store (with and without local caching), S3FS, Perkeep and UtahFS in different Unix mount points and compared them using our four microbenchmarks. S3FS allows Linux and macOS to mount an Amazon S3 bucket via FUSE without any security properties. Perkeep, formerly called Camlistore, is a FUSE-based cryptographic filesystem that can be backed by memory, local or cloud storage. UtahFS which is in its initial stage of development, stores encrypted data on untrusted cloud storage. We mounted UtahFS without Path ORAM that hides the access patterns, as it degrades the performance \cite{utahblospost}. Having the Path ORAM enabled, the \textbf{Write} benchmark runs $18.59 \times$ slower. We configured Perkeep and UtahFS to use an Amazon S3 account for our evaluation. 

\begin{figure}[t!]
	\centering
	\begin{subfigure}{.5\columnwidth}
		\centering
		\fbox{\includegraphics[width=0.9\columnwidth]{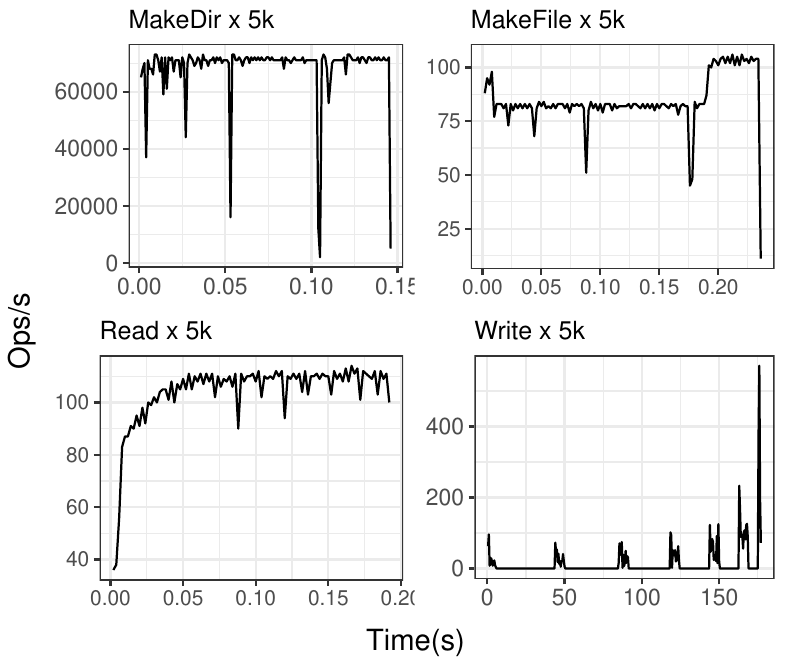}}
		\caption{\fsname{}-global}
		\label{fig:logtime-upss-global}
	\end{subfigure}%
	\begin{subfigure}{.5\columnwidth}
		\centering
		\fbox{\includegraphics[width=0.9\columnwidth]{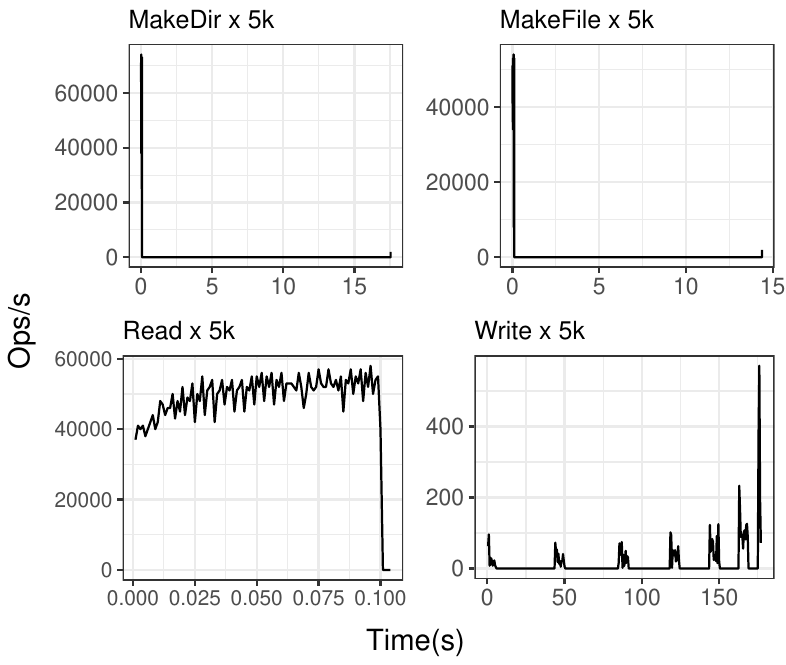}}
		\caption{\fsname{}-global-full-sync}
		\label{fig:logtime-upss-global-persist-once-no-cache}
	\end{subfigure}%
	\\
	\begin{subfigure}{.5\columnwidth}
		\centering
		\fbox{\includegraphics[width=0.9\columnwidth]{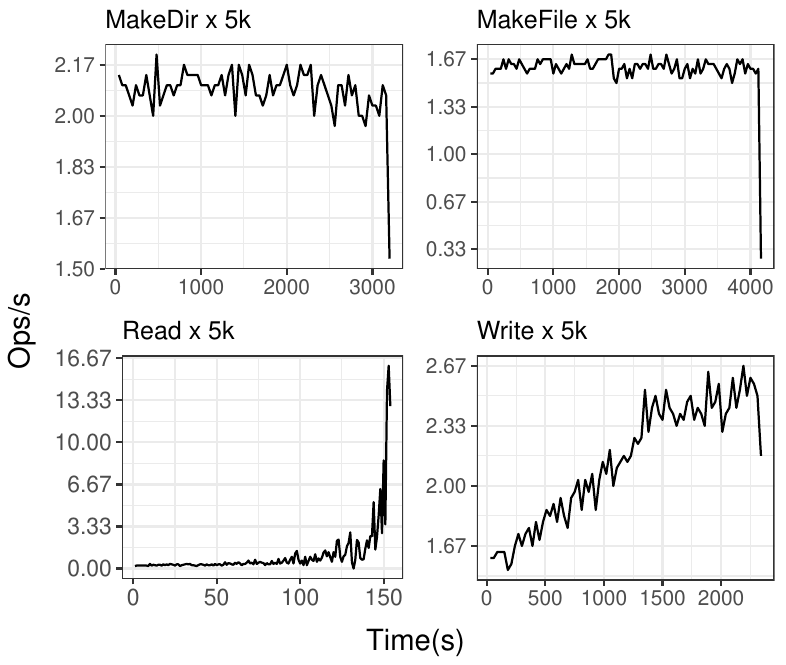}}
		\caption{Perkeep}
		\label{fig:logtime-perkeep}
	\end{subfigure}%
	\begin{subfigure}{0.5\columnwidth}
		\centering
		\fbox{\includegraphics[width=0.9\columnwidth]{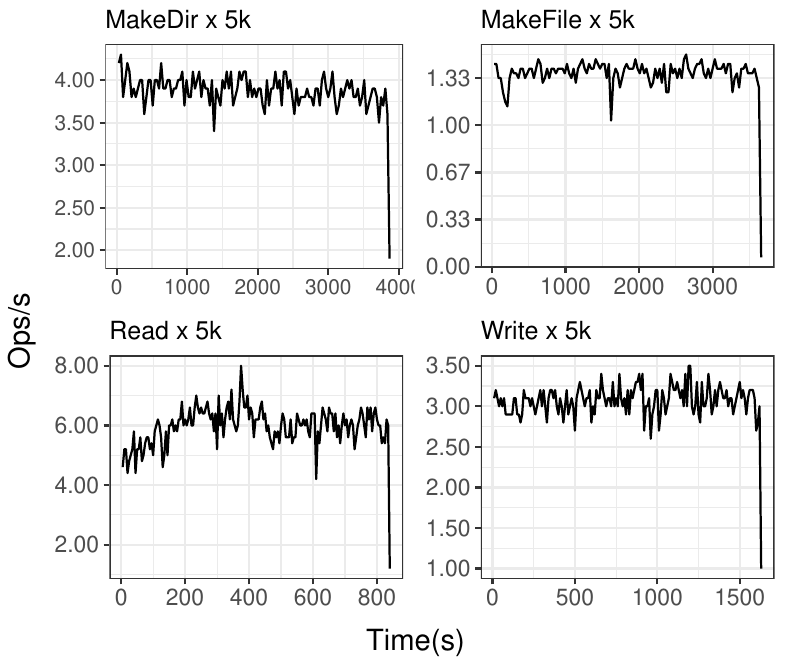}}
		\caption{S3FS}
		\label{fig:logtime-s3fs}
	\end{subfigure}
	\begin{subfigure}{0.5\columnwidth}
		\centering
		\fbox{\includegraphics[width=0.9\columnwidth]{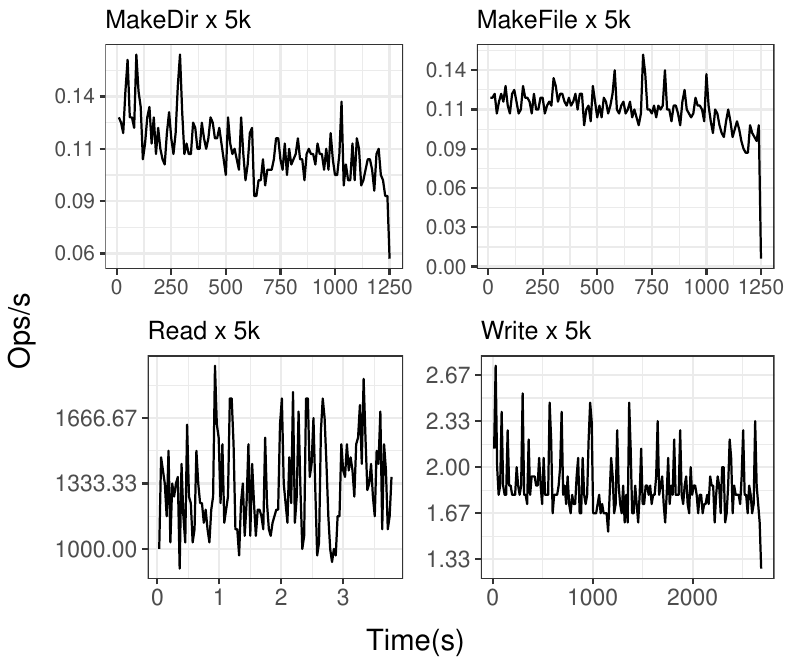}}
		\caption{UtahFS}
		\label{fig:logtime-utahfs}
	\end{subfigure}
	\caption[Operations per second for global filesystems]{
    Performance comparison of \fsname{}-global, S3FS, Perkeep and UtahFS.
    Owing to long read and write delays for comparison filesystems,
    benchmarks were run for 5\,kops rather than 100\,kops.
	  \label{fig:logtime-global}
  }
\end{figure}

We ran the benchmarks discussed in \Cref{sec:local-fs} with 5k \textbf{MakeDir},
\textbf{MakeFile}, \textbf{Read} and \textbf{Write} operations and the
behaviours of \fsname{}-network, S3FS, Perkeep and UtahFS during time are
reported in \Cref{fig:logtime-global}. In all of these cases, Amazon S3's
response time is the bottleneck. To have a fair comparison, we ran the
benchmarks for \fsname{} with and without caching. With caching enabled, we
write the encrypted blocks in a caching block store and journal the blocks to an
on-disk file, then we write to Amazon S3 bucket by processing the journal using
a background thread. This makes a large difference in the number of operations
that can be done by \fsname{} as a global filesystem in comparison with S3FS,
Perkeep and UtahFS (\Cref{fig:logtime-upss-global}). In
\Cref{fig:logtime-upss-global-persist-once-no-cache}, we disabled caching and
persisted the content just before the benchmark script is finished so that the
content is ready to be read from the Amazon block store. Even without caching
and having the content persisted to the Amazon block store, \fsname{}
outperforms the other three filesystems by factors of 10--8,000. These results show that the cryptographic foundation of \systemname{} provides, not just strong security properties, but a foundation for aggressive caching that would be unsafe in a system that does not use cryptographic naming.

\section{\systemname{}: a foundation for novel applications}
\label{sec:applications}

The performance of \systemname{} can be compared to extant filesystems using
\fsname{}, but the most compelling aspects of \systemname{} are in the novel
system designs it can enable.
In this section we describe the possibility of new content-sharing systems
that provide \textit{redaction with integrity} (\Cref{sec:redaction})
and \textit{private-by-default version control} (\Cref{sec:vcs})
based on \systemname{}'s unique characteristics.

\subsection{Redaction with integrity}
\label{sec:redaction}

Organizational settings with strong privacy and security requirements often
necessitate the redaction of documents before sharing or disclosing them.
Redaction is not supported by conventional filesystems, but \systemname{}'s
design allows redaction to be made explicit, and for relationships between
unredacted and redacted versions to be tracked, allowing a digital ``chain of
custody'' even for redacted documents.
By constructing a \rust{Version} for a file containing full block pointers
(block name and key) for some blocks but only block names for others, it is
possible to maintain a full Merkle DAG for blind content.
The \object{prev} pointer in the redacted \object{Version} also contains just
the block name of the original \object{Version}.
Therefore, a user that has access to a redacted file can reference the version
it was derived from but cannot read the unredacted version.
The redacted version can be modified and those changes fed
back to the original while maintaining file or directory integrity.

An \systemname{} object's \rust{redact} method takes two offsets specifying the
range to be redacted and returns a new file or directory object.
\Cref{lst:redact} shows how the second block of a four-block file can be
redacted, with the last block of the redacted file then modified,
and new bytes finally added to the redacted file.
The \rust{diff} method then finds differences between the original and the
redacted file.

\begin{figure}[ht]
\begin{lstlisting}[language=Rust, label=lst:redact, caption={An example of redacting a file and process diff on the original and redacted file.}, captionpos=b]
let f = get_a_file()?;
f.write(&four_blocks)?;

let redacted = f.redact(4096, 4096 + 4095)?;
redacted.set_offset(end - 16)?;
redacted.write("the edited bytes".as_bytes())?;
redacted.write("added bytes".as_bytes())?;
f.diff(redacted)?;
\end{lstlisting}

\begin{lstlisting}[language=diff, label=lst:redact-diff, caption={Output of executing \Cref{lst:redact}.}, captionpos=b]
--- a/file
+++ b/file

@@ -4096,4096 +4096,4096 @@
+++ Redacted

@@ -16368,16 +16368,16 @@
- "JUG47744NENOJPVW"
+ "the edited bytes"

@@ -16384,0 +16384,11 @@
+ "added bytes"
\end{lstlisting}
\end{figure}

\begin{figure}
	\centering
	\includegraphics[width=0.9\columnwidth]{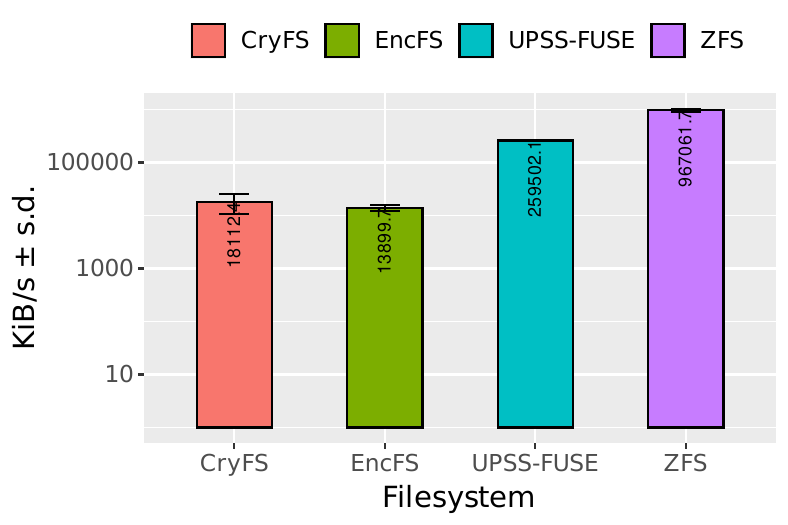}
	\caption[Redaction benchmarks]{
    Redaction performance as compared with CryFS, EncFS and ZFS
    (which zero out rather than redact content).
    Average of 100 runs with standard deviation.
	}
	\label{fig:redaction}
\end{figure}

We evaluated redaction performance by creating 10~MB files and redacting half
of their content starting from a random offset.
We compared \fsname{} performance with a local block store to the filesystems
discussed in \Cref{sec:local-fs} (ZFS, EncFS and CryFS).
Since these filesystems do not support redaction, we simulated redaction in
them by zeroing out bytes in the file, yielding the results in
\Cref{fig:redaction}.
In this figure, \systemname{} outperforms even filesystems with weaker security
properties, although ZFS still shows its maturity and level of optimization
(albeit with no confidentiality enabled).

\subsection{\vcfullname}
\label{sec:vcs}

The construction of \systemname{} with its underlying DAG of immutable blocks
resembles extant filesystems such as ZFS, but also distributed revision control
systems such as Git~\cite{loeliger2012version}.
The Ori filesystem explicitly reduces the file consistency problem to a
version control problem, like \systemname{}~\cite{mashtizadeh2013replication},
but like Git, it does not provide confidentiality guarantees.
\systemname{} provides an opportunity to create a \textit{least-privileged}
revision control system that treats all data as private by default unless
explicitly shared, leveraging \systemname{}'s underlying structure to represent
immutable versions efficiently.
We have begun to prototype such a revision control system, \vcfullname{}.
Its implementation is incomplete --- it does not, for example, authenticate
users or make access control decisions about them.
However, it is complete enough to provide initial performance evaluation that
demonstrates the strong utility of \systemname{} as a basis for such a
revision control system.

In \vcname{}, a client program uses \systemname{} directories to manage a tree
of source code via the \systemname{} API and generate \rust{Version} objects.
Block pointers to these \rust{Version} objects can then be pushed to a
repository that serializes incoming changes from multiple clients into a linear
sequence of repository versions.
New directory versions that are based on the current repository version can
be accepted and treated as the new repository version; directory versions that
are not based on the current version are rejected.
As in Git, such rejected ``push'' operations reveal a need for the client to
pull the current respository version and rebase their work on it.
All clients and the repository server share a remote block store, which
stores encrypted blocks named by cryptographic hashes.

\begin{figure}[ht]
	\centering
	\includegraphics[width=0.9\columnwidth]{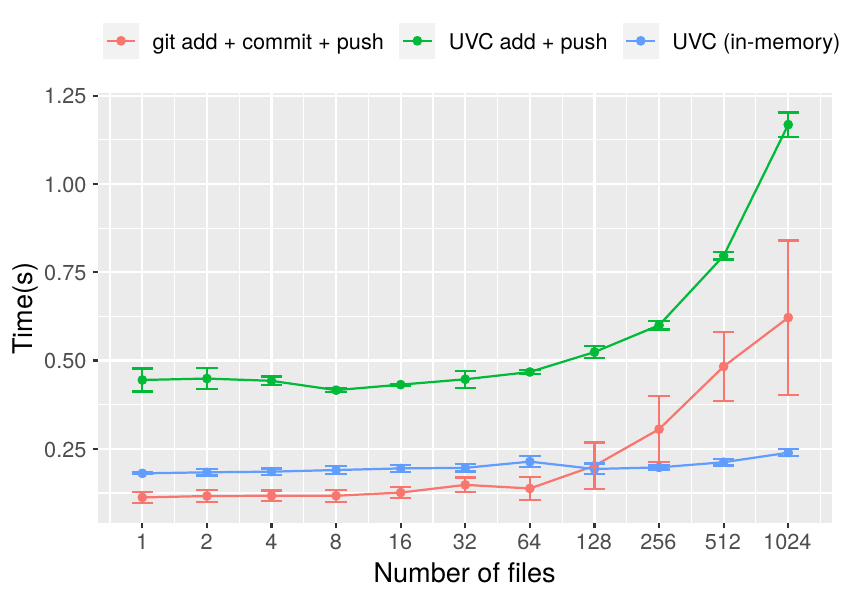}
	\caption[\vcname{} vs. Git - add, commit and push]{
		Time required to add and push files to remote \vcname{} and Git
		repositories.
		Results show the average and standard deviation of five runs.
		\label{fig:total-push}
	}
\end{figure}

For our evaluation, we started with an empty repository and
added a variable number of source files from the Linux kernel to
our repository, using a UVC \command{add} operation or Git's equivalent sequence
\command{add} and \command{commit}.
Finally, we pushed revisions to the remote repository.
The ``remote'' block store was run on the local machine, as was the Git
``remote'', to remove networking costs from our comparative measurements.
This procedure was repeated for increasing number of Linux source files,
up to the first 1,024 files in the Linux kernel source tree, representing
18\,MB of source code.
The time required to complete these operations is shown in
\Cref{fig:total-push}.

\begin{figure}
	\centering
	\includegraphics[width=0.9\columnwidth]{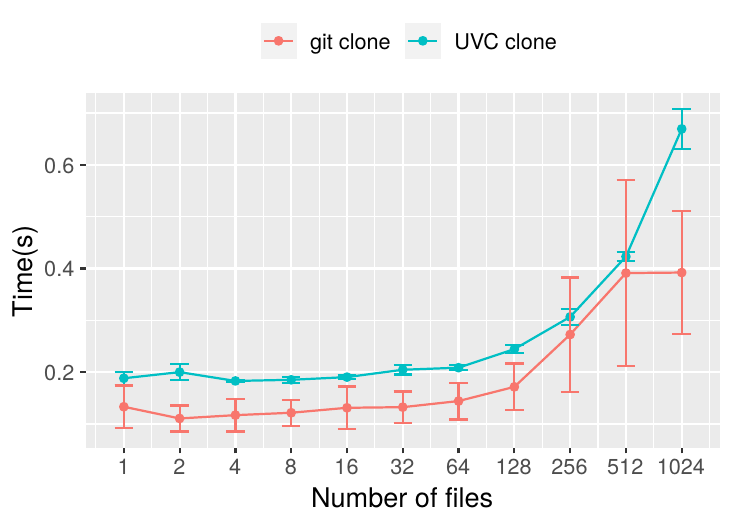}
	\caption[\vcname{} vs. Git - clone]{
		Time required to clone a remote repository in Git and \vcname{}.
		Results show the average and standard deviation of 5 runs.
		\label{fig:clone}
	}
\end{figure}

Despite the additional computational effort required to encrypt all of the data
transferred through \vcname{}, \Cref{fig:total-push} shows that it was less
than 1\,s slower than the mature, intensively-optimised Git for all
measurements, with approximately a 2$\times$ slowdown for a
very large add-and-push operation.
Writing data to the remote block store is the most time-consuming phase of our
push procedure; we anticipate that future multi-threaded communication with the
block store will substantially improve performance.
The ``in-memory'' line in \Cref{fig:total-push} shows the lower bound of
potential cost: preparing commits but not sending blocks to the remote store.
\Cref{fig:clone} shows the time required to clone a remote repository for both
Git and \vcname{}.
In this read-only case, \vcname{}'s performance is substantially closer to that 
of Git, as fewer \systemname{}-specific operations
(persisting, hashing plaintext and hashing ciphertext) need to be performed.

Together, these results demonstrate that \systemname{}'s security model provides
a practical foundation for distributed revision control with far stronger
security properties than today's conventional revision control systems, yet with
performance approaching that of mature, heavily-optimised systems.
\vcname{} is not (yet) a fully-functional replacement for Git, but it
demonstrates the utility of \systemname{}'s approach.

\section{Related work}
\label{sec:ralated-work}

The CFS \cite{dabek2001wide}, Coda \cite{satyanarayanan1990coda},Ivy \cite{muthitacharoen2002ivy}, and FARSITE \cite{adya2002farsite} filesystems provide availability for user data stored on dedicated servers in a distributed environment along with other features such as disconnected operations, content-addressable storage and log-structured systems. Coda introduced an automatic conflict resolution that can detect most but not all the classes of conflicts. Ivy also introduced a conflict detector called \textit{lc} that notifies users about conflicts. Similar to \fsname{}, FARSITE, which is a decentralized network filesystem, uses convergent encryption \cite{douceur2002reclaiming, li2013secure, agarwala2017dice} to protect user data. As CFS, Coda and Ivy are non-cryptographic filesystems,  they cannot rely on untrusted storage servers. On the other hand, the access control lists in FARSITE, which is a cryptographic fileystem, is not completely decoupled from user data; therefore, higher level applications cannot define their own policies, as it is possible in \systemname{}. 

Several filesystems have been designed for untrusted cloud settings, such as
NCryptFS \cite{wright2003ncryptfs}, EncFS \cite{encfs}, OutFS \cite{khashan2020secure} and CryFS
\cite{messmer2017novel}.
NCryptFS and EncFS are cryptographic filesystems, which protect content by
encrypting files, but leave filesystem metadata such as the directory structure
unprotected.
CryFS and OutFS solve this problem by splitting all filesystem data into fixed-size
blocks and encrypting each block individually. CryFs uses one key for all
encryptions, but OutFS generates separate keys per file.
The creators of EncFS expanded their work to make it a multi-user filesystem by
applying Unix local permissions to the encrypted files before being stored on
remote servers \cite{leibenger2016encfs}.
However, EncFS and OutFS are not practical for multi-user environments with
non-local users.

CageCoach \cite{carpenter2023cagecoach} is another distributed and cryptographic 
filesystem that builds on features of \systemname{} that were introduced 
in \cite{upss}. CageCoach makes the first step
towards partial sharing via redaction over encrypted read-only data. 
CageCoach is, as yet, a pre-publication prototype 
under development.

Ori \cite{mashtizadeh2013replication}, IPFS \cite{benet2014ipfs} and Perkeep
\cite{perkeep} (formerly known as Camlistore) connect multiple devices with a
filesystem that users can access anywhere.
IPFS synthesizes key ideas from DHTs \cite{stoica2001chord},
BitTorrent \cite{cohen2003incentives}, Git \cite{loeliger2012version}
and self-certifying pathnames \cite{mazieres1998escaping}
to create a peer-to-peer version-controlled filesystem.
Both Ori and IPFS reduce the data inconsistency problem to a
version control problem by storing new versions of files upon modification;
Ori handles updates with the CoW technique.
Synchronization, failures handling, data recovery and sharing/\textit{grafting}
are key Ori features.
Perkeep uses open protocols to create a unified store for user data from
different sources such as Twitter or a local hard drive.
Similar to \fsname{}, Perkeep can be backed by a memory store, a local store
or a cloud account.
However, none of Ori, IPFS or Perkeep provide a mechanism for
sharing redacted file and directory hierarchies. Moreover, Perkeep leaves the directory structure unprotected on the backing service. 

MetaSync \cite{han2015metasync} and DepSky \cite{bessani2013depsky} are
synchronization services that store confidential data on untrusted cloud storage
providers.
MetaSync synchronizes multiple cloud storage providers using pPaxos and a
deterministic replication algorithm to maintain a globally consistent view of
the synchronized files.
DepSky provides availability, integrity and confidentiality of data stored on
four different cloud storage servers, replicated by quorum techniques.
It uses symmetric-key encryption and distributes the key between clouds using a
secret sharing scheme, so no individual cloud service can recover the key alone. However, these two systems cannot be used as a platform for novel applications that \systemname{} can support and they just synchronize multiple cloud services.

Tahoe \cite{wilcox2008tahoe} and UtahFS \cite{utahfs} are cryptographic filesystems with the goal of storing user data on untrusted storage servers.
As in \systemname{}, Tahoe and UtahFS store content encrypted in Merkle DAGs and provide access control by cryptographic capabilities.
Unlike \systemname{}, however, Tahoe's replica-oriented design lends itself more readily to storage of shared immutable data than to the use cases of a general-purpose filesystem.
In Tahoe, files are encrypted with a symmetric key, the ciphertext is erasure-coded using Reed-Solomon codes \cite{rizzo1997feasibility} and split into $N$ shares to be written to $N$ servers.
Mutable files are signed and verified with a public/private key pair, which is stored alongside the file.
Mutation of the file requires knowledge of this private key, and mutation by multiple collaborators can cause data inconsistency.
Furthermore, mutation requires copying and encrypting entire files, which does
not lend itself to the frequent mutation that is found in practical filesystems.
By contrast, \systemname{}'s blocks, block pointers and \object{Version} structure allow arbitrarily-small ranges of files to be mutated frequently without overly-zealous copying or re-encryption.
UtahFS encrypts files with one symmetric key; files larger than a predefined size are broken into fixed-size blocks.
UtahFS supports hiding access patterns by Path ORAM \cite{stefanov2013path}, but this feature is disabled by default as it degrades performance significantly \cite{utahblospost}.

\section{Future work}

\subsection{FFI}
\label{sec:future-ffi}
Currently, applications integrating the \systemname{} library access it via
Rust linkage and calling conventions~\cite{rust}.
However, \systemname{} supports compilation into WebAssembly~\cite{wasm};
this will allow us to explore Web-based experiences in which user data is
decrypted within a user's browser only.
In the future we will also support other programming languages such as C and
Python via foreign function interfaces.

\subsection{Structured files. }
\label{sec:crdt}
Files in classical filesystems are unstructured byte arrays.
However, the internal DAG structure of \systemname{} blocks should allow \systemname{} to naturally define structured files to better represent complex data without serialization or deserialization~\cite{ahmed2012file}.
In this way, multi-user data on different replicas are guaranteed to be in the same state, without data loss and without requiring users to resolve conflicts manually. Automatic filesystem-level conflict resolution has been explored before in filesystems such as Coda \cite{satyanarayanan1990coda}, but \systemname{}' internal block structure naturally lends itself to a reinvigorated exploration of these ideas, defining files as Conflict-free Replicated Data Types (CRDT) \cite{kleppmann2017conflict, shapiro2011conflict, shapiro2011comprehensive}. 

\subsection{Blind auditing}
\label{sec:blind-auditing}

Systems in security- and privacy-conscious settings require the ability to
audit accesses to data, ideally without opening a large attack surface by
exposing all plaintext data to auditors.
Treating backend storage as a sea of encrypted blocks provides an opportunity
for auditing accesses to data without revealing that data to auditors:
users might authenticate to a future block store to allow their access patterns
to be observed without revealing private data.
In such an environment, authorization systems could produce sets of
permissible-to-access blocks to be compared with actual block accesses,
and ``honeypot'' records could be monitored by block hash without revealing
their plaintext.

\section{Conclusion}
\label{sec:conclusion}

\textit{\systemfullname{}} provides data availability, strong confidentiality
and integrity properties while relying only on untrusted backend storage (local or remote).
Data is encrypted at rest, named cryptographically and store within a
content-addressable \textit{sea of blocks}, so
no file or directory structure can be discerned directly from the contents
of an encrypted block store.
Cryptographic capabilities are used to authorize access to arbitrarily-sized
DAGs of files and directories without centralized access control.
Convergent encryption enables data de-duplication for large files among even
mutually-distrustful users while avoiding the common pitfalls of the technique
for small, low-entropy files.

\systemname{} wraps copy-on-write operations with a conventional filesystem API,
accessible directly as a library or proxied via a FUSE interface.
Although \fsname{}'s performance is lower than that of direct API usage,
it exceeds that of
comparable cryptographic filesystems and is within an order of magnitude of
that of the mature copy-on-write filesystem ZFS.
When using remote storage, \systemname{}'s performance exceeds that of
UtahFS, Google's Perkeep and even, for some benchmarks, unencrypted NFS.

Beyond performance comparison with conventional filesystems, we have also
demonstrated that \systemname{}'s design provides useful primitives for building
novel privacy- and security-conscious applications.
Specifically, we have demonstrated that \systemname{} can be used to build
systems that support \textit{redaction with integrity} as well as
\textit{least-privileged revision control}.
Such systems would be prohibitively expensive to build without the unique
features afforded by \systemname{}.

\systemname{} demonstrates that it is possible to achieve both
strong security properties \textit{and} high performance, backed by untrusted
local, remote or global storage.
\systemname{}'s performance is comparable to
--- or, in some cases, superior to --- mature, heavily-optimized filesystems.
Adoption of \systemname{} will lay the foundation for
future transformations in privacy and integrity for applications as diverse as
social networking and medical data storage, providing better opportunities for
users --- not system administrators --- to take control of their data.

\bibliographystyle{plain}
\bibliography{upssfs}

\begin{thebibliography}{10}

\bibitem{filebench}
{Filebench - A model based filesystem workload generator}.
\newblock \url{https://github.com/filebench/filebench}, July 2016.

\bibitem{wasm}
{WebAssembly Specification}.
\newblock \url{https://webassembly.org}, 2017.

\bibitem{fuse}
{FUSE (Filesystem in Userspace)}.
\newblock \url{ https://github.com/libfuse/libfuse/releases/tag/fuse-3.9.0},
  Dec 2019.

\bibitem{utahfs}
{UtahFS}.
\newblock \url{https://github.com/cloudflare/utahfs/releases/tag/v1.0}, June
  2020.

\bibitem{utahblospost}
{UtahFS: Encrypted File Storage}.
\newblock \url{https://blog.cloudflare.com/utahfs}, June 2020.

\bibitem{adya2002farsite}
Atul Adya, William~J Bolosky, Miguel Castro, Gerald Cermak, Ronnie Chaiken,
  John~R Douceur, Jon Howell, Jacob~R Lorch, Marvin Theimer, and Roger~P
  Wattenhofer.
\newblock {FARSITE}: Federated, available, and reliable storage for an
  incompletely trusted environment.
\newblock {\em ACM SIGOPS Operating Systems Review}, 36(SI):1--14, 2002.

\bibitem{agarwala2017dice}
Ashish Agarwala, Priyanka Singh, and Pradeep~K Atrey.
\newblock {DICE: A dual integrity convergent encryption protocol for client
  side secure data deduplication}.
\newblock In {\em {2017 IEEE International Conference on Systems, Man, and
  Cybernetics (SMC)}}, pages 2176--2181. IEEE, 2017.

\bibitem{ahmed2012file}
Mehdi Ahmed-Nacer, St{\'e}phane Martin, and Pascal Urso.
\newblock {File system on CRDT}.
\newblock {\em arXiv preprint arXiv:1207.5990}, 2012.

\bibitem{amazons3}
{Amazon Web Services, Inc.}
\newblock {Amazon Simple Storage Service}.
\newblock \url"https://aws.amazon.com/s3", (Accessed on February 28, 2020).

\bibitem{Asklund:1999:extensional-versioning}
Ulf Asklund, Lars Bendix, Henrik~B. Christensen, and Boris Magnusson.
\newblock The unified extensional versioning model.
\newblock In {\em System Configuration Management}, pages 100--122, Berlin,
  Heidelberg, 1999. Springer Berlin Heidelberg.

\bibitem{baker1991measurements}
Mary~G Baker, John~H Hartman, Michael~D Kupfer, Ken~W Shirriff, and John~K
  Ousterhout.
\newblock {Measurements of a distributed file system}.
\newblock In {\em {Proceedings of the thirteenth ACM Symposium on Operating
  Systems Principles}}, pages 198--212, 1991.

\bibitem{benet2014ipfs}
Juan Benet.
\newblock {{IPFS}}: content addressed, versioned, {{P2P}} file system.
\newblock {\em arXiv preprint arXiv:1407.3561}, 2014.

\bibitem{bessani2013depsky}
Alysson Bessani, Miguel Correia, Bruno Quaresma, Fernando Andr{\' e}, and Paulo
  Sousa.
\newblock {DepSky}: dependable and secure storage in a cloud-of-clouds.
\newblock {\em ACM Transactions on Storage (TOS)}, 9(4):1--33, 2013.

\bibitem{bonwick2003zettabyte}
Jeff Bonwick, Matt Ahrens, Val Henson, Mark Maybee, and Mark Shellenbaum.
\newblock {The Zettabyte file system}.
\newblock In {\em {Proc. of the 2nd Usenix Conference on File and Storage
  Technologies}}, volume 215, 2003.

\bibitem{bozorgi2020online}
Arastoo Bozorgi.
\newblock {\em {From online social network analysis to a user-centric private
  sharing system}}.
\newblock PhD thesis, Memorial University of Newfoundland, 2020.

\bibitem{icissp24}
Arastoo Bozorgi., Mahya Jadidi., and Jonathan Anderson.
\newblock Upss: A global, least-privileged storage system with stronger
  security and better performance.
\newblock In {\em Proceedings of the 10th International Conference on
  Information Systems Security and Privacy - ICISSP}, pages 660--671. INSTICC,
  SciTePress, 2024.

\bibitem{upss}
Arastoo Bozorgi, Mahya~Soleimani Jadidi, and Jonathan Anderson.
\newblock {Challenges in Designing a Distributed Cryptographic File System}.
\newblock In {\em {Cambridge International Workshop on Security Protocols}},
  pages 177--192. Springer, 2019.

\bibitem{carpenter2023cagecoach}
Jason Carpenter and Zhi-Li Zhang.
\newblock Cagecoach: Sharing-oriented redaction-capable distributed
  cryptographic file system.
\newblock {\em arXiv preprint arXiv:2301.04214}, 2023.

\bibitem{cohen2003incentives}
Bram Cohen.
\newblock Incentives build robustness in {{BitTorrent}}.
\newblock In {\em {Workshop on Economics of Peer-to-Peer systems}}, volume~6,
  pages 68--72, 2003.

\bibitem{dabek2001wide}
Frank Dabek, M~Frans Kaashoek, David Karger, Robert Morris, and Ion Stoica.
\newblock {Wide-area cooperative storage with CFS}.
\newblock In {\em {ACM SIGOPS Operating Systems Review}}, volume~35, pages
  202--215. ACM, 2001.

\bibitem{capabilitysystems}
J.~B. Dennis and E.~C. Van~Horn.
\newblock {Programming semantics for multiprogrammed computations}.
\newblock {\em Communications of the ACM}, 9(3):143--155, 1966.

\bibitem{douceur2002reclaiming}
John~R Douceur, Atul Adya, William~J Bolosky, P~Simon, and Marvin Theimer.
\newblock {Reclaiming space from duplicate files in a serverless distributed
  file system}.
\newblock In {\em {Proceedings of the 22nd International Conference on
  Distributed Computing Systems}}, pages 617--624. IEEE, 2002.

\bibitem{dworkin2015sha}
Morris Dworkin.
\newblock {SHA-3 Standard: Permutation-Based Hash and Extendable-Output
  Functions}.
\newblock {Federal Inf. Process. Stds. (NIST FIPS)}, National Institute of
  Standards and Technology, 2015.

\bibitem{nist:2001:aes}
Morris~J. Dworkin, Elaine~B. Barker, James~R. Nechvatal, James Foti,
  Lawrence~E. Bassham, E.~Roback, and James F.~Drawy Jr.
\newblock {Advanced Encryption Standard (AES)}.
\newblock {Federal Inf. Process. Stds. (NIST FIPS)}, National Institute of
  Standards and Technology, 2001.

\bibitem{s3fs}
Andrew Gaul, Takeshi Nakatani, and rrizun.
\newblock {S3FS: FUSE-based file system backed by Amazon S3)}.
\newblock \url{https://github.com/s3fs-fuse/s3fs-fuse/releases}, Feb 2020.

\bibitem{han2015metasync}
Seungyeop Han, Haichen Shen, Taesoo Kim, Arvind Krishnamurthy, Thomas Anderson,
  and David Wetherall.
\newblock {MetaSync}: File synchronization across multiple untrusted storage
  services.
\newblock In {\em 2015 USENIX Annual Technical Conference (USENIX ATC 15)},
  pages 83--95, 2015.

\bibitem{rfc2898}
Burt Kaliski.
\newblock {PKCS \#5: Password-Based Cryptography Specification Version 2.0}.
\newblock RFC 2898, 2000.

\bibitem{khashan2020secure}
Osama~Ahmed Khashan.
\newblock Secure outsourcing and sharing of cloud data using a user-side
  encrypted file system.
\newblock {\em IEEE Access}, 8:210855--210867, 2020.

\bibitem{kleppmann2017conflict}
Martin Kleppmann and Alastair~R Beresford.
\newblock A conflict-free replicated {{JSON}} datatype.
\newblock {\em IEEE Transactions on Parallel and Distributed Systems},
  28(10):2733--2746, 2017.

\bibitem{lazowska1986file}
Edward~D Lazowska, John Zahorjan, David~R Cheriton, and Willy Zwaenepoel.
\newblock {File access performance of diskless workstations}.
\newblock {\em ACM Transactions on Computer Systems (TOCS)}, 4(3):238--268,
  1986.

\bibitem{leibenger2016encfs}
Dominik Leibenger, Jonas Fortmann, and Christoph Sorge.
\newblock {EncFS goes multi-user: Adding access control to an encrypted file
  system}.
\newblock In {\em {2016 IEEE Conference on Communications and Network Security
  (CNS)}}, pages 525--533. IEEE, 2016.

\bibitem{li2013secure}
Jin Li, Xiaofeng Chen, Mingqiang Li, Jingwei Li, Patrick~PC Lee, and Wenjing
  Lou.
\newblock {Secure deduplication with efficient and reliable convergent key
  management}.
\newblock {\em IEEE Transactions on Parallel and Distributed Systems},
  25(6):1615--1625, 2013.

\bibitem{perkeep}
Paul Lindner and Wil Norris.
\newblock {Perkeep (n\'ee Camlistore): your personal storage system for life.}
\newblock \url{https://github.com/perkeep/perkeep/releases}, May 2018.

\bibitem{loeliger2012version}
Jon Loeliger and Matthew McCullough.
\newblock {\em {Version Control with Git: Powerful tools and techniques for
  collaborative software development}}.
\newblock O'Reilly Media, Inc., 2012.

\bibitem{mashtizadeh2013replication}
Ali~Jos{\'e} Mashtizadeh, Andrea Bittau, Yifeng~Frank Huang, and David
  Mazieres.
\newblock Replication, history, and grafting in the {{Ori}} file system.
\newblock In {\em {Proceedings of the Twenty-Fourth ACM Symposium on Operating
  Systems Principles}}, pages 151--166. ACM, 2013.

\bibitem{mazieres1998escaping}
David Mazieres and M~Frans Kaashoek.
\newblock {Escaping the evils of centralized control with self-certifying
  pathnames}.
\newblock In {\em {Proceedings of the 8th ACM SIGOPS European workshop on
  Support for composing distributed applications}}, pages 118--125. ACM, 1998.

\bibitem{merkle:1979:thesis}
R.~Merkle.
\newblock {\em {Secrecy, authentication, and public key systems}}.
\newblock PhD thesis, 1979.

\bibitem{messmer2017novel}
Sebastian Messmer, Jochen Rill, Dirk Achenbach, and J{\"o}rn M{\"u}~ller Quade.
\newblock {A novel cryptographic framework for cloud file systems and CryFS, a
  provably-secure construction}.
\newblock In {\em {IFIP Annual Conference on Data and Applications Security and
  Privacy}}, pages 409--429. Springer, 2017.

\bibitem{azure}
{Microsoft, Inc.}
\newblock {Azure Blob Storage}.
\newblock \url"https://azure.microsoft.com/en-us/products/storage/blobs/ ",
  (Accessed on January, 2023).

\bibitem{muthitacharoen2002ivy}
Athicha Muthitacharoen, Robert Morris, Thomer~M Gil, and Benjie Chen.
\newblock {Ivy: A read/write peer-to-peer file system}.
\newblock {\em ACM SIGOPS Operating Systems Review}, 36(SI):31--44, 2002.

\bibitem{rizzo1997feasibility}
Luigi Rizzo.
\newblock {On the feasibility of software FEC}.
\newblock {\em Univ. di Pisa, Italy}, pages 1--16, 1997.

\bibitem{rosenblum1992design}
Mendel Rosenblum and John~K Ousterhout.
\newblock {The design and implementation of a log-structured file system}.
\newblock {\em ACM Transactions on Computer Systems (TOCS)}, 10(1):26--52,
  1992.

\bibitem{satyanarayanan1990coda}
Mahadev Satyanarayanan, James~J. Kistler, Puneet Kumar, Maria~E. Okasaki,
  Ellen~H. Siegel, and David~C. Steere.
\newblock {Coda: A highly available file system for a distributed workstation
  environment}.
\newblock {\em IEEE Transactions on Computers}, 39(4):447--459, 1990.

\bibitem{shapiro2011comprehensive}
Marc Shapiro, Nuno Pregui{\c{c}}a, Carlos Baquero, and Marek Zawirski.
\newblock {\em {A comprehensive study of convergent and commutative replicated
  data types}}.
\newblock PhD thesis, Inria--Centre Paris-Rocquencourt; INRIA, 2011.

\bibitem{shapiro2011conflict}
Marc Shapiro, Nuno Pregui{\c{c}}a, Carlos Baquero, and Marek Zawirski.
\newblock {Conflict-free replicated data types}.
\newblock In {\em {Symposium on Self-Stabilizing Systems}}, pages 386--400.
  Springer, 2011.

\bibitem{nfs}
S.~Shepler, B.~Callaghan, D.~Robinson, R.~Thurlow, C.~Beame, M.~Eisler, and
  D.~Noveck.
\newblock {RFC3530: Network File System (NFS) Version 4 Protocol}, 2003.

\bibitem{stefanov2013path}
Emil Stefanov, Marten Van~Dijk, Elaine Shi, Christopher Fletcher, Ling Ren,
  Xiangyao Yu, and Srinivas Devadas.
\newblock {Path ORAM: an extremely simple oblivious RAM protocol}.
\newblock In {\em {Proceedings of the 2013 ACM SIGSAC conference on Computer \&
  communications security}}, pages 299--310, 2013.

\bibitem{stoica2001chord}
Ion Stoica, Robert Morris, David Karger, M~Frans Kaashoek, and Hari
  Balakrishnan.
\newblock {Chord: A scalable peer-to-peer lookup service for internet
  applications}.
\newblock {\em ACM SIGCOMM Computer Communication Review}, 31(4):149--160,
  2001.

\bibitem{encfs}
EncFS Team.
\newblock {EncFS: an Encrypted Filesystem for FUSE}.
\newblock \url{https://github.com/vgough/encfs/releases/tag/v1.9.5}, 2018.

\bibitem{rust}
Rust Team.
\newblock {Rust programming language}.
\newblock \url{https://www.rust-lang.org}, 2020.

\bibitem{sshfs}
SSHFS Team.
\newblock {{SSHFS}} (a network filesystem client to connect to ssh servers).
\newblock \url{https://github.com/libfuse/sshfs/releases}, Jan 2020.

\bibitem{wilcox2008tahoe}
Zooko Wilcox-O'Hearn and Brian Warner.
\newblock {Tahoe: the least-authority filesystem}.
\newblock In {\em {Proceedings of the 4th ACM International Workshop on Storage
  Security and Survivability}}, pages 21--26, 2008.

\bibitem{wright2003ncryptfs}
Charles~P Wright, Michael~C Martino, and Erez Zadok.
\newblock {NCryptfs: A Secure and Convenient Cryptographic File System.}
\newblock In {\em {USENIX Annual Technical Conference, General Track}}, pages
  197--210, 2003.

\end{thebibliography}

\end{document}